\documentclass[twocolumn,superscriptaddress,showpacs,preprintnumbers,amsmath,amssymb]{revtex4}
\newcommand{\Bx}{x_{\rm B}}

\newcommand{\pEpg}{$\vec ep~\rightarrow~ep\gamma$}
\newcommand{\Epg}{$ep~\rightarrow~ep\gamma$}

\newcommand{\EpX}{$ep~\rightarrow~epX$}
\newcommand{\pEpX}{$\vec ep~\rightarrow~epX$}

\begin{document}


\title{{\large\bf Beam Spin Asymmetries in DVCS with CLAS} at $4.8$ GeV}

\newcommand*{\UNH}{University of New Hampshire, Durham, New Hampshire 03824-3568}
\affiliation{\UNH}
\newcommand*{\ODU}{Old Dominion University, Norfolk, Virginia 23529}
\affiliation{\ODU}
\newcommand*{\JLAB}{Thomas Jefferson National Accelerator Facility, Newport News, Virginia 23606}
\affiliation{\JLAB}
\newcommand*{\ANL}{Argonne National Laboratory}
\affiliation{\ANL}
\newcommand*{\ASU}{Arizona State University, Tempe, Arizona 85287-1504}
\affiliation{\ASU}
\newcommand*{\UCLA}{University of California at Los Angeles, Los Angeles, California  90095-1547}
\affiliation{\UCLA}
\newcommand*{\CSU}{California State University, Dominguez Hills, Carson, CA 90747}
\affiliation{\CSU}
\newcommand*{\CMU}{Carnegie Mellon University, Pittsburgh, Pennsylvania 15213}
\affiliation{\CMU}
\newcommand*{\CUA}{Catholic University of America, Washington, D.C. 20064}
\affiliation{\CUA}
\newcommand*{\SACLAY}{CEA-Saclay, Service de Physique Nucl\'eaire, 91191 Gif-sur-Yvette, France}
\affiliation{\SACLAY}
\newcommand*{\CNU}{Christopher Newport University, Newport News, Virginia 23606}
\affiliation{\CNU}
\newcommand*{\UCONN}{University of Connecticut, Storrs, Connecticut 06269}
\affiliation{\UCONN}
\newcommand*{\ECOSSEE}{Edinburgh University, Edinburgh EH9 3JZ, United Kingdom}
\affiliation{\ECOSSEE}
\newcommand*{\FU}{Fairfield University, Fairfield CT 06824}
\affiliation{\FU}
\newcommand*{\FIU}{Florida International University, Miami, Florida 33199}
\affiliation{\FIU}
\newcommand*{\FSU}{Florida State University, Tallahassee, Florida 32306}
\affiliation{\FSU}
\newcommand*{\GWU}{The George Washington University, Washington, DC 20052}
\affiliation{\GWU}
\newcommand*{\ECOSSEG}{University of Glasgow, Glasgow G12 8QQ, United Kingdom}
\affiliation{\ECOSSEG}
\newcommand*{\ISU}{Idaho State University, Pocatello, Idaho 83209}
\affiliation{\ISU}
\newcommand*{\INFNFR}{INFN, Laboratori Nazionali di Frascati, 00044 Frascati, Italy}
\affiliation{\INFNFR}
\newcommand*{\INFNGE}{INFN, Sezione di Genova, 16146 Genova, Italy}
\affiliation{\INFNGE}
\newcommand*{\ORSAY}{Institut de Physique Nucleaire ORSAY, Orsay, France}
\affiliation{\ORSAY}
\newcommand*{\ITEP}{Institute of Theoretical and Experimental Physics, Moscow, 117259, Russia}
\affiliation{\ITEP}
\newcommand*{\JMU}{James Madison University, Harrisonburg, Virginia 22807}
\affiliation{\JMU}
\newcommand*{\KYUNGPOOK}{Kyungpook National University, Daegu 702-701, Republic of Korea}
\affiliation{\KYUNGPOOK}
\newcommand*{\MIT}{Massachusetts Institute of Technology, Cambridge, Massachusetts  02139-4307}
\affiliation{\MIT}
\newcommand*{\UMASS}{University of Massachusetts, Amherst, Massachusetts  01003}
\affiliation{\UMASS}
\newcommand*{\MOSCOW}{Moscow State University, General Nuclear Physics Institute, 119899 Moscow, Russia}
\affiliation{\MOSCOW}
\newcommand*{\NSU}{Norfolk State University, Norfolk, Virginia 23504}
\affiliation{\NSU}
\newcommand*{\OHIOU}{Ohio University, Athens, Ohio  45701}
\affiliation{\OHIOU}
\newcommand*{\PITT}{University of Pittsburgh, Pittsburgh, Pennsylvania 15260}
\affiliation{\PITT}
\newcommand*{\RPI}{Rensselaer Polytechnic Institute, Troy, New York 12180-3590}
\affiliation{\RPI}
\newcommand*{\RICE}{Rice University, Houston, Texas 77005-1892}
\affiliation{\RICE}
\newcommand*{\URICH}{University of Richmond, Richmond, Virginia 23173}
\affiliation{\URICH}
\newcommand*{\SCAROLINA}{University of South Carolina, Columbia, South Carolina 29208}
\affiliation{\SCAROLINA}
\newcommand*{\UNIONC}{Union College, Schenectady, NY 12308}
\affiliation{\UNIONC}
\newcommand*{\VT}{Virginia Polytechnic Institute and State University, Blacksburg, Virginia   24061-0435}
\affiliation{\VT}
\newcommand*{\VIRGINIA}{University of Virginia, Charlottesville, Virginia 22901}
\affiliation{\VIRGINIA}
\newcommand*{\WM}{College of William and Mary, Williamsburg, Virginia 23187-8795}
\affiliation{\WM}
\newcommand*{\YEREVAN}{Yerevan Physics Institute, 375036 Yerevan, Armenia}
\affiliation{\YEREVAN}

\newcommand*{\NONE}{unknown}
\newcommand*{\deceased}{Deceased}
\newcommand*{\NOWOHIOU}{Ohio University, Athens, Ohio  45701}
\newcommand*{\NOWINDSTRA}{Systems Planning and Analysis, Alexandria, Virginia 22311}
\newcommand*{\NOWUNH}{University of New Hampshire, Durham, New Hampshire 03824-3568}
\newcommand*{\NOWMOSCOW}{Moscow State University, General Nuclear Physics Institute, 119899 Moscow, Russia}
\newcommand*{\NOWMIT}{Massachusetts Institute of Technology, Cambridge, Massachusetts  02139-4307}
\newcommand*{\NOWECOSSEE}{Edinburgh University, Edinburgh EH9 3JZ, United Kingdom}
\newcommand*{\NOWGEISSEN}{Physikalisches Institut der Universitaet Giessen, 35392 Giessen, Germany}
\newcommand*{\NOWPASCAL}{Universit\'e Blaise Pascal/ CNR-IN2P3 63177
Aubi\`ere, France}
\author {G.~Gavalian} 
\affiliation{\UNH}
\affiliation{\ODU}
\author {V.D.~Burkert} 
\affiliation{\JLAB}
\author {L.~Elouadrhiri} 
\affiliation{\JLAB}
\author {M.~Holtrop} 
\affiliation{\UNH}
\author {S.~Stepanyan} 
\affiliation{\JLAB}
\author {D.~Abrahamyan} 
\affiliation{\ISU}
\author {G.~Adams} 
\affiliation{\RPI}
\author {M.J.~Amaryan} 
\affiliation{\ODU}
\author {P.~Ambrozewicz} 
\affiliation{\FIU}
\author {M.~Anghinolfi} 
\affiliation{\INFNGE}
% ...
\collaboration{The CLAS Collaboration}
\noaffiliation

\date{\today}

\begin{abstract}             

We report measurements of the beam spin asymmetry in Deeply
Virtual Compton Scattering (DVCS) at an electron beam energy of $4.8$ GeV using
the CLAS detector at the Thomas Jefferson National Accelerator Facility. 
The DVCS beam spin asymmetry has been measured in a wide range of
kinematics, $1.0$ (GeV/c)$^2$ $<Q^2<2.8$ (GeV/c)$^2$, $0.12<x_B<0.48$, and $0.1$
(GeV/c)$^2$ $<-t<0.8$ (GeV/c)$^2$, using the reaction \pEpX. 
The number of H$(e,e^{\prime}\gamma p)$ and H$(e,e^{\prime}\pi^0 p)$
events are separated in each $(Q^2,x_B,t)$ bin by a fit to the
line shape of the H$(e,e^{\prime}p)X$ $M_x^2$ distribution.
The validity of the method was
studied in detail using experimental and simulated data. It was
shown that with the achieved missing mass squared resolution and the
available statistics, the separation of DVCS-BH and $\pi^0$
events can reliably be done with less than 5\% uncertainty.
The $Q^2$- and $t$-dependences of the 
$\sin\phi$ moments of the asymmetry are extracted and
compared with theoretical calculations. 


\end{abstract}

\maketitle

\centerline{PACS numbers: 13.60.z, 14.20.Dh, 24.85.+p}

\section{Introduction}

\indent

Hard scattering processes play an important role in the understanding of
the quark and gluon structure of hadrons. The important feature of
hard reactions is the possibility of separating the perturbative (short
distance) and non-perturbative (long distance) parts of the
interaction. 
This so-called factorization
property has been successfully used in inclusive measurements ({\it
  e.g.} in Deep Inelastic Scattering 
(DIS) of leptons) to study the internal structure of the nucleon.
Until recently, very
few exclusive  processes could be treated in the framework of perturbative QCD (pQCD) and
compared to experimental data  (typical examples are the
$\pi^0\gamma\gamma^*$ transition form factor \cite{j_gron} and the elastic form
factors of the pion \cite{piff} and the nucleon \cite{belnff}).  
The recently developed formalism of a QCD description of Deeply
Virtual Compton Scattering (DVCS) \cite{Ji97,Rady} and Deeply
Exclusive Meson Production \cite{CFS}, provides a
framework in which 
the amplitudes of these processes can be factorized into a
hard-scattering part  
(exactly calculable in pQCD) and a non-perturbative nucleon  structure 
part that can be parameterized at the amplitude level by means of
Generalized Parton Distributions (GPDs). The GPDs contain information
on quark/antiquark correlations, particularly the correlation of their
transverse spatial and longitudinal momentum distributions, and on the
quark angular momentum \cite{mburk1}. They provide a unifying picture
for an entire set of fundamental quantities containing information
on the hadronic structure,
such as nucleon form factors (which are related to matrix elements of
vector and
axial vector currents), polarized and unpolarized parton
distributions, and contributions to the spin of the nucleon due to
orbital excitations of quarks and gluons.  

There are four chiral-even GPDs, denoted
$H^q$, $\tilde H^q$, $E^q$, and $\tilde E^q$, which depend on the kinematical
variables $x$, $\xi$, and $t$. 
They correspond to the amplitude for removing a quark with momentum
fraction $x+\xi$ and restoring it with momentum fraction $x-\xi$. 
The light-cone momentum
fraction $x$ is defined by $k^+ = x \bar{P}^+$, where $k$ is the quark loop
momentum and $\bar{P}$ is the average nucleon momentum
($\bar{P}=(p^\prime + p)/2$, where $p$ and $p^\prime$ are the initial
and final state nucleon four-momenta, respectively). $\xi$ is the
generalized Bjorken variable, $\xi=Q^2/(4q\cdot \bar{P}) \rightarrow x_B/(2-x_B)$ as $t/Q^2 \rightarrow 0$, where
$q=k-k^\prime$ and $Q^2=-q^2$. $k$ and  
$k^\prime$ are the initial and final electron momenta. 

The Mandelstam variable $t=\Delta^{2}=(p^\prime -p)^{2}$ is the Lorentz-invariant four momentum transfer squared to the target. 
In the forward limit, $t\to 0$, the GPDs $H$ and  
$\tilde{H}$ reduce to the quark density distributions $q(x)$ and quark
helicity distributions $\Delta q(x)$ obtained from DIS:

\begin{eqnarray}
\label{eq:dislimit}
H^{q}(x,0,0)\,
&=& \left\{
\begin{array}{cr}
q(x),& \hspace{.5cm} x \; > \; 0\, \\
- \bar q(-x),& \hspace{.5cm} x \; < \; 0 \,
\end{array}
\right. \\
\hspace {0.5cm}
\label{eq:dislimitp}
\tilde{H}^{q}(x,0,0)\,
&=& \left\{
\begin{array}{cr}
\Delta q(x),& \hspace{.5cm} x \; > \; 0\, \\
\Delta \bar q(-x),& \hspace{.5cm} x \; < \; 0 \, .
\end{array}
\right.
\end{eqnarray}

$E$ and $\tilde{E}$ are accessible through hard exclusive
electro-production reactions only and are new leading-twist functions.
Similarly, in DIS, which corresponds to the limit $\xi \rightarrow 0$,
the region $-\xi < x < \xi$ is absent. In this region the GPDs behave like
meson distribution amplitudes and contain completely new information
about nucleon structure.
At finite momentum transfer, the first moments of the GPDs are related to
the elastic form factors of the nucleon through model-independent sum
rules. By integrating over $x$ one can obtain for a particular quark
flavor (for any $\xi$):
\begin{equation}
\int_{-1}^{+1} dx H^{q}(x,\xi,t) = F_{1}^{q}(t)
\end{equation}
\begin{equation}
\int_{-1}^{+1} dx E^{q}(x,\xi,t) = F_{2}^{q}(t)
\end{equation}
\begin{equation}
\int_{-1}^{+1} dx \tilde{H}^{q}(x,\xi,t) = g_{A}^{q}(t)
\end{equation}
\begin{equation}
\int_{-1}^{+1} dx \tilde{E}^{q}(x,\xi,t) = h_{A}^{q}(t) ,
\end{equation}
where $F_{1}^{q}(t)$ and $F_{2}^{q}(t)$ represent the elastic Dirac
and Pauli form factors, respectively, for the quark flavor $q$ in the nucleon,
$g_{A}^{q}$ is the axial-vector form factor of the nucleon, and
$h_{A}^{q}$ is the pseudo-scalar form factor.

\begin{figure}[ht]
\vspace{30mm} 
{\includegraphics{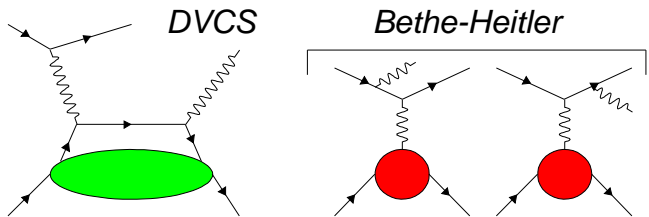}}
\caption
{(Color) Feynman diagrams for DVCS and Bethe-Heitler processes
contributing to the amplitude of \Epg~ scattering.}
\label{fig:DVCS_BH_FDIA}
\end{figure}

Deeply Virtual Compton Scattering (DVCS), $ep\rightarrow ep\gamma$, is the
simplest reaction to access GPDs experimentally. 
DVCS has the same final state as the Bethe-Heitler (BH) process, 
where a photon is emitted by the incoming or outgoing electron
(Fig.~\ref{fig:DVCS_BH_FDIA}). 
The differential cross section for the process $ep\rightarrow ep\gamma$
can be written as \cite{bel0112}:
\begin{equation}
{d\sigma \over{dx_Bdydtd\phi}}={\alpha^3x_By
\over{8\pi Q^2\sqrt{1+\epsilon^2}}}|{\cal T}|^2.
\label{eq:cs}
\end{equation}
Here the amplitude of the production of a real photon, $\cal
T$, is the sum of the DVCS (${\cal T}_{DVCS}$) and BH
(${\cal T}_{BH}$) amplitudes, given as:
\begin{equation}
{\cal T}^2 = |{\cal T}_{BH}|^2 + |{\cal T}_{DVCS}|^2 + {\cal I},
\label{eq:arph}
\end{equation}
where 
\begin{equation}
{\cal I} = {\cal T}_{DVCS}{\cal T}^*_{BH} + {\cal T}^*_{DVCS}{\cal T}_{BH}
\end{equation}
is the interference term. ${\cal T}_{BH}$ is real to the lowest order
in the QED fine structure constant.  The lepton energy fraction  $y$ and Bjorken variable $x_B$ are defined as:
\begin{eqnarray}
y={p\cdot q \over {p\cdot k}},~~ x_B={Q^2 \over {2p\cdot q}}.
\end{eqnarray}
In the notation
of Ref.~\cite{bel0112} $\epsilon=2x_BM/Q$, where $M$ is the nucleon mass.
$\phi$ is the angle between the leptonic plane and the proton-photon
production plane.
%

Contraction of the leptonic and hadronic
tensors generates an azimuthal angular dependence of each of the three terms in
Eq.(\ref{eq:arph}) \cite{diehl}. In a frame with the z-axis along the
virtual photon, the dependence of the amplitudes on $\phi$ yields a finite sum of Fourier 
harmonics. The amplitude of deeply virtual production of a photon has been
derived up to twist-3 accuracy \cite{bel0112}. In the notation of
Ref.~\cite{bel0112}, the helicity-dependent angular moments are presented in a
series of $\sin(n\phi)$, with $n=1,2,3$.
Only $n=1$ is a twist-2 quark matrix element. The $n=2$ terms
are twist-3 and the $n=3$ terms are twist-2 double-helicity-flip
gluon transversity terms, which are kinematically suppressed. If these terms are omitted, then at the twist-2 level, the BH, DVCS, and interference
contributions to the total cross section in Eq.(\ref{eq:arph}) read

\begin{widetext}
\begin{eqnarray}
\label{Par-BH}
&&{|{\cal T}_{\rm BH}|^2}
= \frac{1}
{\Bx^2 y^2 (1 + \epsilon^2)^2 \Delta^2\, {\cal P}_1 (\phi) {\cal P}_2 (\phi)}
\left\{
c^{\rm BH}_0 + c^{\rm BH}_1 \, \cos{\phi} + c^{\rm BH}_2 \, \cos{2\phi}
\right\} \, ,
\\
\label{AmplitudesSquared}
&& |{\cal T}_{\rm DVCS}|^2
=
\frac{1}{y^2 Q^2}\left\{
c^{\rm DVCS}_0
\right\} \, ,
\\
\label{InterferenceTerm}
&& {\cal I}
= \frac{-1}{\Bx y^3 \Delta^2 {\cal P}_1 (\phi) {\cal P}_2 (\phi)}
\left\{
c_0^{\cal I} +
c_1^{\cal I} \cos\phi +  s_1^{\cal I} \sin\phi
\right\} \, ,
\end{eqnarray}
\end{widetext}
where ${\cal P}_1 (\phi)$ and ${\cal P}_2 (\phi)$ are the BH
propagators.

Experimentally, the simplest observable to measure GPDs
is the beam spin asymmetry ($A_{LU}$).  The largest contribution
to this observable arises from  the imaginary part of the
interference of the DVCS and the BH amplitudes, $s_1^{\cal I}$.
$s_1^{\cal I}$ is the most interesting Fourier harmonic since it is
linear in the Compton Form Factor (CFF) \cite{bel0112n}

\begin{eqnarray}
s_1^{\cal I} = 8K\lambda y(2-y)\Im{\rm m}\left[ C^I({\cal F})\right]\, ,
\label{eq:intxt14}
\end{eqnarray}
where $K$ is the $\sqrt{-t/Q^2}$ power-suppressed kinematical factor and $\lambda$ is the beam polarization.
Thus, at the twist-2 level, the
helicity-dependent cross-section difference will be:

\begin{widetext}
\begin{equation}
A_{LU}(\phi)\cdot(\sigma^+ + \sigma^- ) = \sigma^+ - \sigma^- 
= \frac{-16Ky(2-y)} {\Bx y^3 \Delta^2 {\cal P}_1 (\phi) {\cal P}_2 (\phi)}(F_1{\cal H} + 
\frac{x_B}{2-x_B} (F_1+F_2) {\tilde {\cal H}} - \frac{t}{4M^2}F_2{\cal E}
)\sin{\phi} .
\label{eq:sfm}
\end{equation}
\end{widetext}

Measuring the imaginary part of the CFF $[{\cal H},{\tilde{\cal H}},{\cal E}]$ gives access to the
GPDs $[H,~{\tilde {H}},~E]$ at the
specific kinematical point $x=\xi$.
From a phenomenological point of view, the extraction of the GPDs
from data requires an extensive
experimental program and detailed analysis with controlled
theoretical corrections.
This field is rapidly expanding (see e.g.
Refs.~\cite{gpv0106,deihl,mburk2}), and as a first 
step, the phenomenological parameterization of GPDs is used
to fit the experimental data.

The azimuthal dependence of the beam spin asymmetry has been 
measured by the CLAS~\cite{Clas01a} and HERMES~\cite{Her01a}
Collaborations with electron ($4.25$~GeV) and positron ($27.6$ GeV)
beams, respectively. The longitudinal target-spin asymmetry has been
reported by CLAS \cite{shifeng}. 
The $Q^2$ dependence of the
helicity-dependent and helicity-independent cross sections was
measured by the Hall A Collaboration at Jefferson Lab \cite{halla}.
The beam charge asymmetry \cite{FR_01} and transverse target spin asymmetry \cite{FR_02} have also been reported by HERMES Collaboration.
Recently, CLAS Collaboration published
measurements of the DVCS beam spin asymmetry over a wide kinematical
range using high statistics data obtained with a 5.77 GeV longitudinally
polarized electron beam \cite{dvcshallb}.
In this paper, new results on the beam spin asymmetry in DVCS at $4.8$
GeV using the CLAS detector are presented. 
Measuring the same observables at the same ($x_B$, $Q^2$, t) but
different energies, will yield different combinations of the terms of interest,
and thus contribute to the determination of GPDs.
In this analysis the number of single photon
events has been extracted using the method developed in Ref.~\cite{Clas01a}. 
A fit to the line
shape of the missing mass squared distributions in the reaction $\vec{e}p\to
epX$ was employed in each kinematical bin. The $Q^2$ and $t$ dependences
of the $\sin{\phi}$ moments have been extracted and are compared with theoretical
predictions.

\section{EXPERIMENT}\label{experiment}

\indent

The measurement was carried out using the CLAS detector \cite{clas} in
Hall B at the Thomas Jefferson National Accelerator Facility. 
The CLAS detector system is based on a toroidal magnet that consists of six
superconducting coils. The coils are positioned symmetrically around the
beam line and provide a toroidal field in the azimuthal
direction. Each of the
gaps between the coils is instrumented with an identical detector
package covering typically $80\%$ of $2\pi$ in the azimuthal direction. 

The CLAS detector package includes
three regions of Drift Chambers (DC) and a set of Scintillator
Counters (SC) for the tracking and identification of long-lived  
charged particles using the time-of-flight technique. The DC and SC
cover the laboratory polar angular 
range from $8^o$ to $140^o$. 
In the forward region, CLAS has gas threshold \v{C}erenkov Counters
(CC) and Electromagnetic Calorimeters (EC) for electron
identification. The electromagnetic calorimeters are also used for detection and identification of photons and neutrons. 
The momentum resolution of the
system is approximately $1\%$. The efficiency of single charged particle detection is
$\ge 95\%$ in the fiducial volume of the detector.

The data used in this analysis were taken during February-March of
2000. The $4.8$ GeV 
longitudinally polarized electron beam was
incident on a $5$-cm long liquid-hydrogen target. The average beam
current was $5$ nA. The total integrated luminosity was $1.28$
fb$^{-1}$.  
The CLAS Data Acquisition system (DAQ) was triggered by a coincidence
of signals from the \v{C}erenkov Counters and the Electromagnetic
Calorimeters. The DAQ rate was $1.1$ kHz at $95\%$ live
time. 
The beam polarization was measured several times
during the run using the Hall B Moller polarimeter. The average 
polarization was $70\%$ with an uncertainty of $\pm 3\%$. 

A total of $1.26$ M triggers have been processed, of 
which about $18\%$ contained a properly identified electron. For
the physics analysis, events with only one detected electron and one
detected proton
were used. For systematic checks, the final states ($ep\gamma$) and
($ep\gamma\gamma$) were analyzed.

\section{Particle Identification}

\indent

At the initial stage of data processing, the three-momenta of all final state
particles are defined and preliminary particle identification (PID)
is performed. Later, in the physics analysis, these PIDs are refined
using the knowledge of the event kinematics. 

\begin{figure}[ht!]
\vspace{90mm} 
{\includegraphics{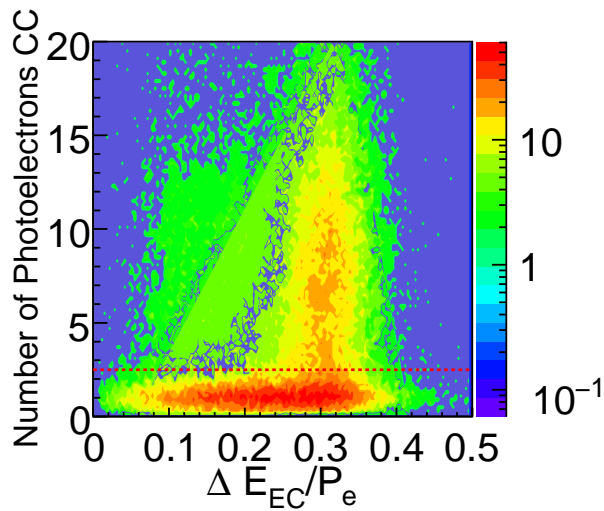}}
\caption
{(Color) Distribution of the number of photoelectrons detected in the CC vs. the
sampling fraction of the EC.}
\label{fig:eccc}
\end{figure}

For the identification of electrons, the energy deposited in the EC and the
number of photoelectrons detected in the CC are used. 
This method relies on the correct reconstruction of the shower energy
and on the analysis of the electromagnetic shower profile. Fiducial cuts in
the calorimeter plane are used to reject particles that
pass close to the edges and lose part of their energy outside of the
calorimeter detection volume. In Fig. \ref{fig:eccc} the distribution
of the number of photoelectrons detected in the CC, $N_{p.e.}$, vs. the
ratio $\Delta E_{EC}/P$ is shown after the
fiducial cuts have been applied. Here $\Delta E_{EC}$ is the
electromagnetic shower energy detected in the EC and $P$ is the momentum
of the particle reconstructed in the DC. The lower horizontal band corresponds to
negative pions. The vertical band, centered at $\Delta E_{EC}/P \simeq
0.27$ (average electromagnetic shower energy sampling fraction of the EC),
corresponds to electrons. A cut $N_{p.e.}>2.5$ is used to reject most
of the pions. Then, the distribution $\Delta E_{EC}/P$ is fitted by a
Gaussian function and a $\pm 3\sigma$ cut around the mean is
imposed to further clean the electron sample.

\begin{figure}[ht]
\vspace{100mm} 
{\includegraphics{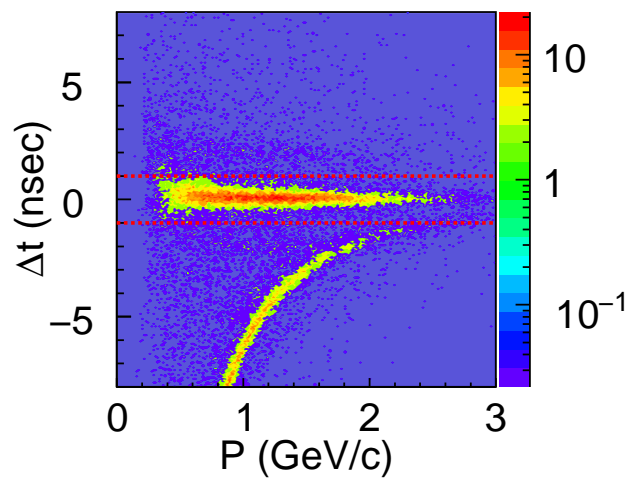}}
\caption
{(Color) Vertex time distribution as a function of momentum for positive
tracks. The particle velocity is deduced from the momentum using
the proton mass.}
\label{fig:vtime}
\end{figure}

Momentum and time-of-flight (TOF)
analysis are performed to identify protons. The CLAS TOF system allows the
separation of protons from kaons up to $2$ GeV/c and protons from pions
up to $3$ GeV/c momentum. In the kinematics of this analysis, protons have
momenta smaller than $1$ GeV/c. Protons are selected using a cut on the 
vertex time, $\Delta t$, calculated by:
\begin{equation}
\label{eq:dt}
\Delta t=t-R/\beta-t_s .
\end{equation}
Here, $t$ is the time measured in the SC and $R$ is the track length from the
production vertex to the scintillator plane, which is determined by the
tracking routines. In Eq. (\ref{eq:dt}), $\beta=p/\sqrt{p^2+m^2}$, where
$p$ and $m$ are the 
proton momentum and rest mass. The event start time, $t_s$, is defined by the electron in
the event as $t_s=t_e-R_e/c$, where $t_e$ is the time measured in the SC
for the electron track and $R_e$ is the electron track length from the
production vertex to the scintillator plane.
In Fig. \ref{fig:vtime}, the dependence of $\Delta t$ on the momentum
is presented for positively charged particles assuming they are
protons. The horizontal band centered at $0$ ns 
corresponds to protons. The lower band, emerging from negative $\Delta t$'s
and closing to the proton band at high momentum, corresponds to
positive pions. A  $\pm 1$ ns cut, shown by the dashed lines, is used to select
protons. 
 
\begin{figure}[ht]
\vspace{90mm} 
{\includegraphics{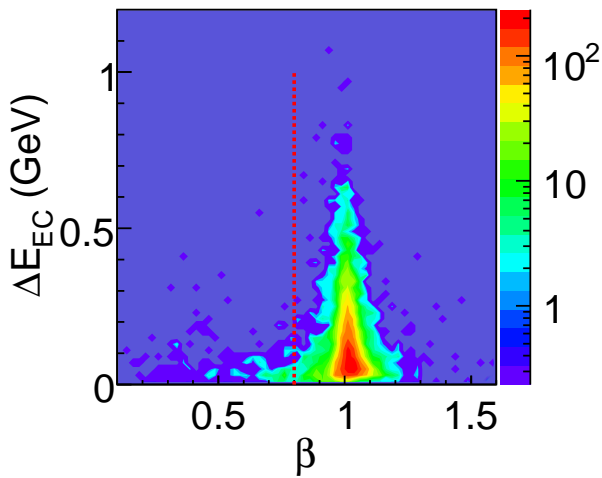}}
\caption
{(Color) Distribution of the energy measured in the EC
vs. $\beta$ for neutral particles detected in the EC. The vertical band at
$\beta=1$ corresponds to photons.}
\label{fig:betaec}
\end{figure}

While final beam spin asymmetries for DVCS-BH events were obtained
using fits to the missing mass distributions in the reaction $ep\to
epX$, 
events with $ep\gamma$ and
$ep\gamma\gamma$ are used to extract the fit parameters and to study
systematic uncertainties. 
The neutral particles in CLAS are detected and 
identified in the electromagnetic calorimeter. 
Neutrons and photons in the EC 
are separated by analyzing the speed of the neutral hits
($\beta_{EC}$). For the selection of photons, a cut on $\beta_{EC}>0.85$
is used, see Fig. \ref{fig:betaec}.  
The energy of the photon is reconstructed using the energy deposited
in the EC, corrected for the sampling fraction of the calorimeter of $0.27$.

\section{Kinematic Corrections}

\indent

Two types of corrections have been applied to the measured momenta of
electrons and protons. The first one is a
correction to the proton momentum to account for energy loss in the
material of the target and in CLAS. The second one is a
 correction to account for small uncertainties in
the magnetic field map and in the drift chamber alignment. 

The energy loss corrections for protons are derived using the GEANT simulation package for
the CLAS detector. Protons are simulated in the kinematic region covered by the
experimental data. The ratio of the generated over the reconstructed 
momenta is parameterized as a function of the reconstructed momentum using
a polynomial function. This function is then used to correct
the measured momentum in the analysis of the experimental data.  

To correct for the effects of the DC misalignments and the uncertainties
in the magnetic field distribution, two methods, applied to
kinematically complete reactions, were studied. The first
method corrects only the magnitude of the electron and proton
momenta. In the second method, a complex fitting algorithm was used to
derive corrections for momenta 
and angles of all charged particles in an exclusive event. These
corrections depend on the momentum, angle, and  charge of the final
state particle and do not depend on the particle type. 

\begin{figure}[ht]
\vspace{160mm} 
{\includegraphics{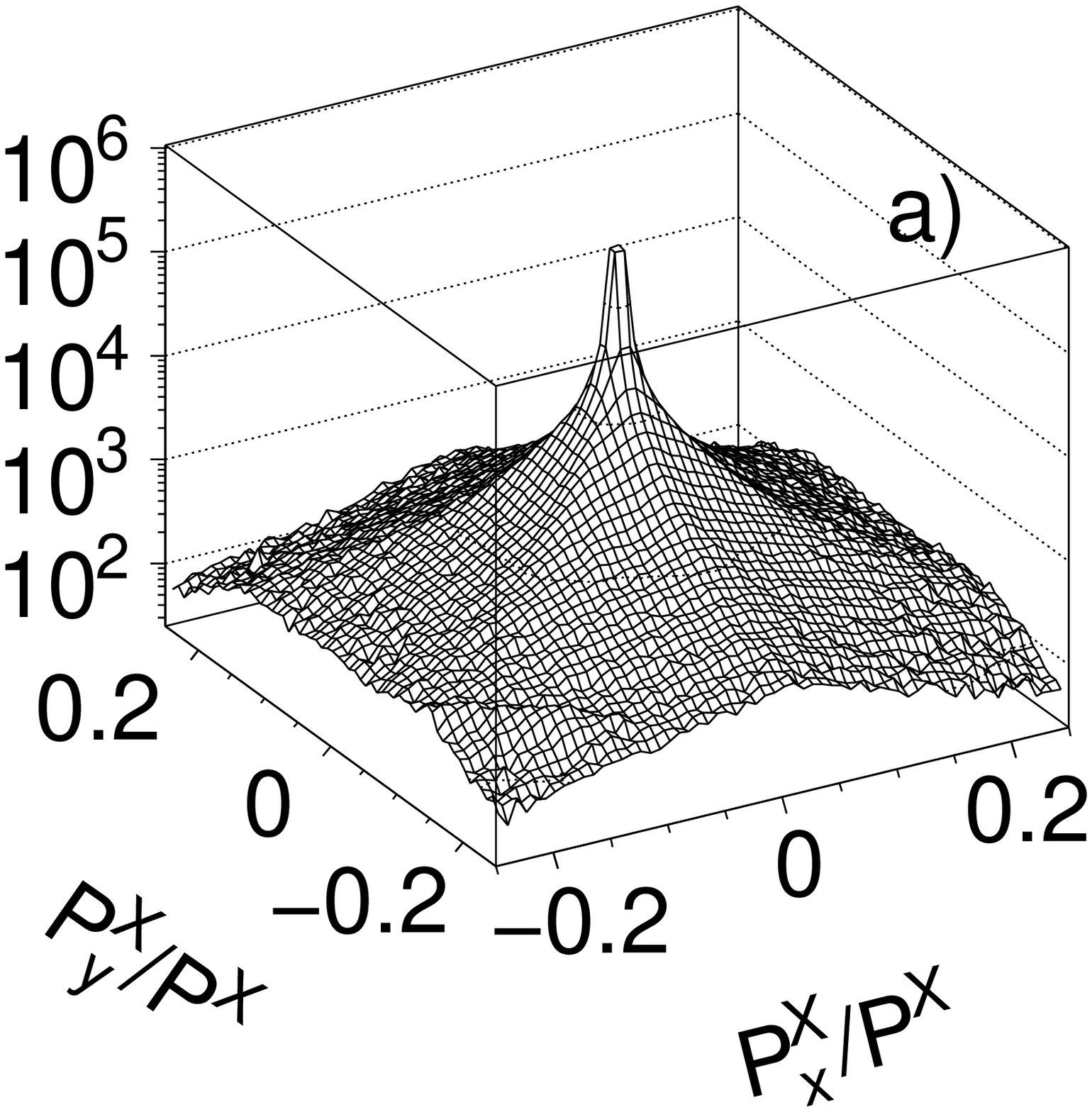}}
{\includegraphics{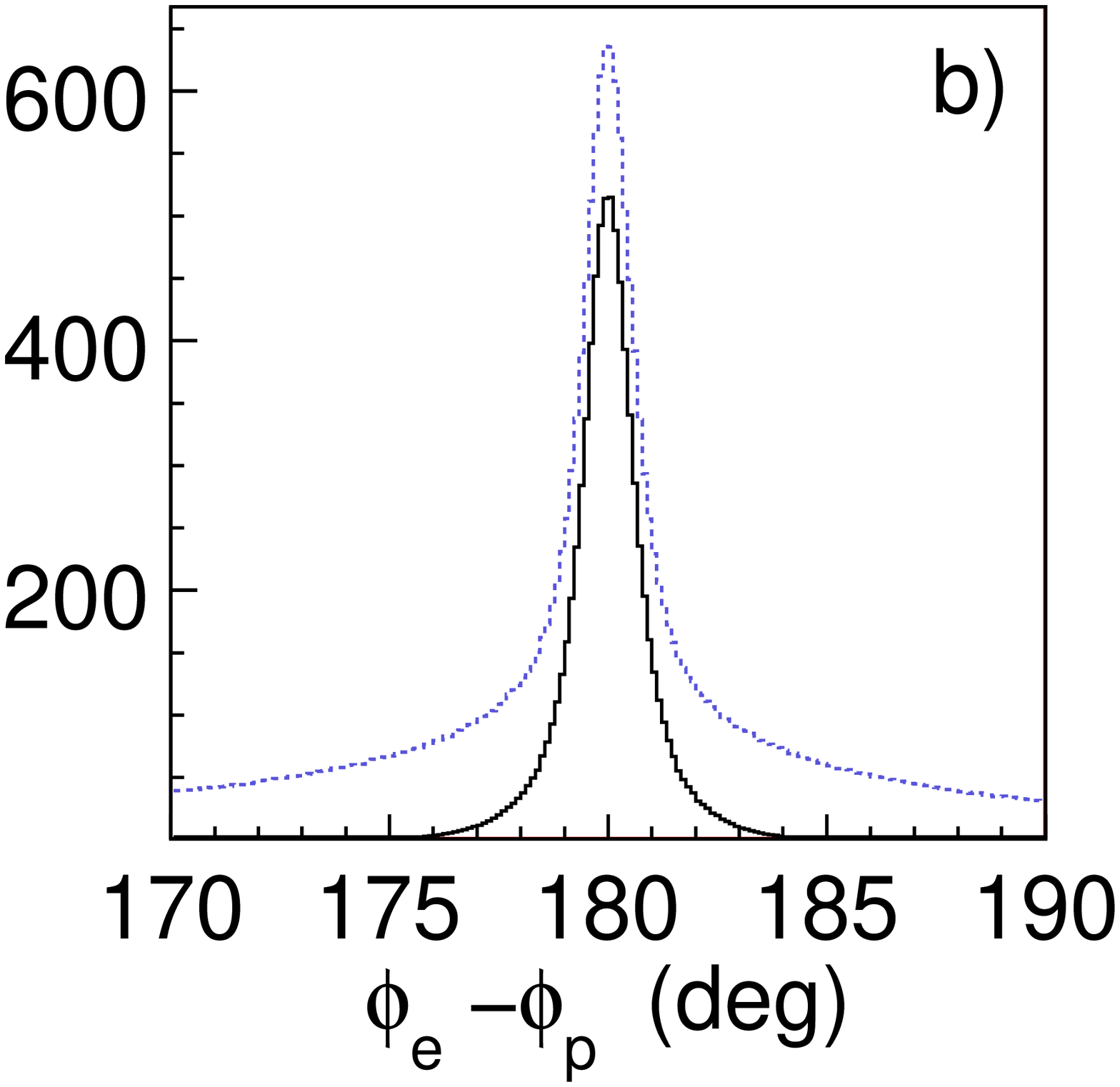}}
\caption
{(Color) Selection of the BH events. a) The distribution of
the $x$ and $y$ components of the missing momentum. The strong peak
at zero corresponds to BH events. b) The electron and proton azimuthal angle
difference before (dashed line histogram) and after (solid
line histogram) the cut on the $x$ and $y$ components of the missing
momentum.}
\label{fig:pxpy}
\end{figure}

In the first method, the electron momentum corrections are derived using
the Bethe-Heitler (BH) events associated with radiation of an
energetic photon by the incoming electron. The kinematics of the
secondary electrons in such events are similar to the 
kinematics in the deep inelastic scattering regime. For the selection of 
the BH events, cuts on the missing mass squared, on the transverse
component of the missing momentum, and on the difference of azimuthal
angles of the electron ($\phi_e$) and the proton ($\phi_p$) in the
reaction $ep\to epX$ are used. In 
Fig.~\ref{fig:pxpy} a) the missing transverse momentum distribution is
shown for events with the missing mass squared within $\pm 0.1$
(GeV/c$^2$)$^2$. The corresponding distribution for the azimuthal angle
difference, $|\phi_e-\phi_p|$, is shown in Fig. \ref{fig:pxpy} b) with
a dashed line histogram. A cut $\sqrt{(P^x_m)^2+(P^y_m)^2}/P_m
< 0.01$ has been 
applied to select events with missing momentum in the direction of
the beam (the radiated photon is in the direction of the incoming
electron). Here $P_m$ is the magnitude and
$P^x_m$ and $P^y_m$ are the transverse components of the missing
momentum. In Fig.~\ref{fig:pxpy} b), the solid line histogram corresponds
to the azimuthal angle difference after the cut on the transverse
component of the missing momentum. The distribution peaks at $180^\circ$,
as it should for the
elastic scattering events. A cut $178^\circ<|\phi_e-\phi_p|<182^\circ$ was
used to select the final event sample. 

The scattered electron momentum,  $P_f^e$, was 
calculated using the measured polar angles:
\begin{eqnarray}
P_f^e &=& \frac{P_i^e}{1+\frac{P_i^e}{M} (1-\cos{\theta_e}) } \nonumber \\
P_i^e &=& \frac{M}{1-\cos{\theta_e}}(\cos{\theta_e} + \cos{\theta_p}
\frac{\sin\theta_e}{\sin{\theta_p}} - 1) ,
\end{eqnarray}    
where $\theta_e$ and $\theta_p$ are the polar angles of the electron and
the proton, respectively, and $M$ is the proton mass. $P_i^e$ is the energy
of the interacting electron after radiation. The ratio of $P_f^e$ over the
 measured momentum was parameterized as a function of the measured angles
and momenta, and used as a correction factor
to the measured electron momentum.

For the proton momentum, corrections derived for the $\pi^+$
from the reaction $ep\to en\pi^+$ with
a missing neutron are used. 
The 
$e\pi^+n$ events are selected using a $\pm 3\sigma$ cut around the neutron
mass peak. The pion momentum was calculated after applying corrections to
the electron momentum and from the assumption that the
missing particle is a neutron. As above, the ratio of the calculated and
measured momenta was parameterized as a function of the measured angles
and momenta, and was used as a correction factor.


In the second method, we used the Bethe-Heitler and $\pi^0$ production
events. The BH events are selected in the same way as
above. In order to identify the pion events, two photons, in addition
to an electron and a proton, have to be detected. A cut on the
invariant mass of the two photons was used for the final
selection of $\pi^0$ events.
Using the missing mass squared of the photon and $\pi^0$ as a constraint in
the fit, corrections for the momenta and polar angles of the electron and
the proton are derived.

\begin{figure}[ht!]
\vspace{180mm} 
{\includegraphics{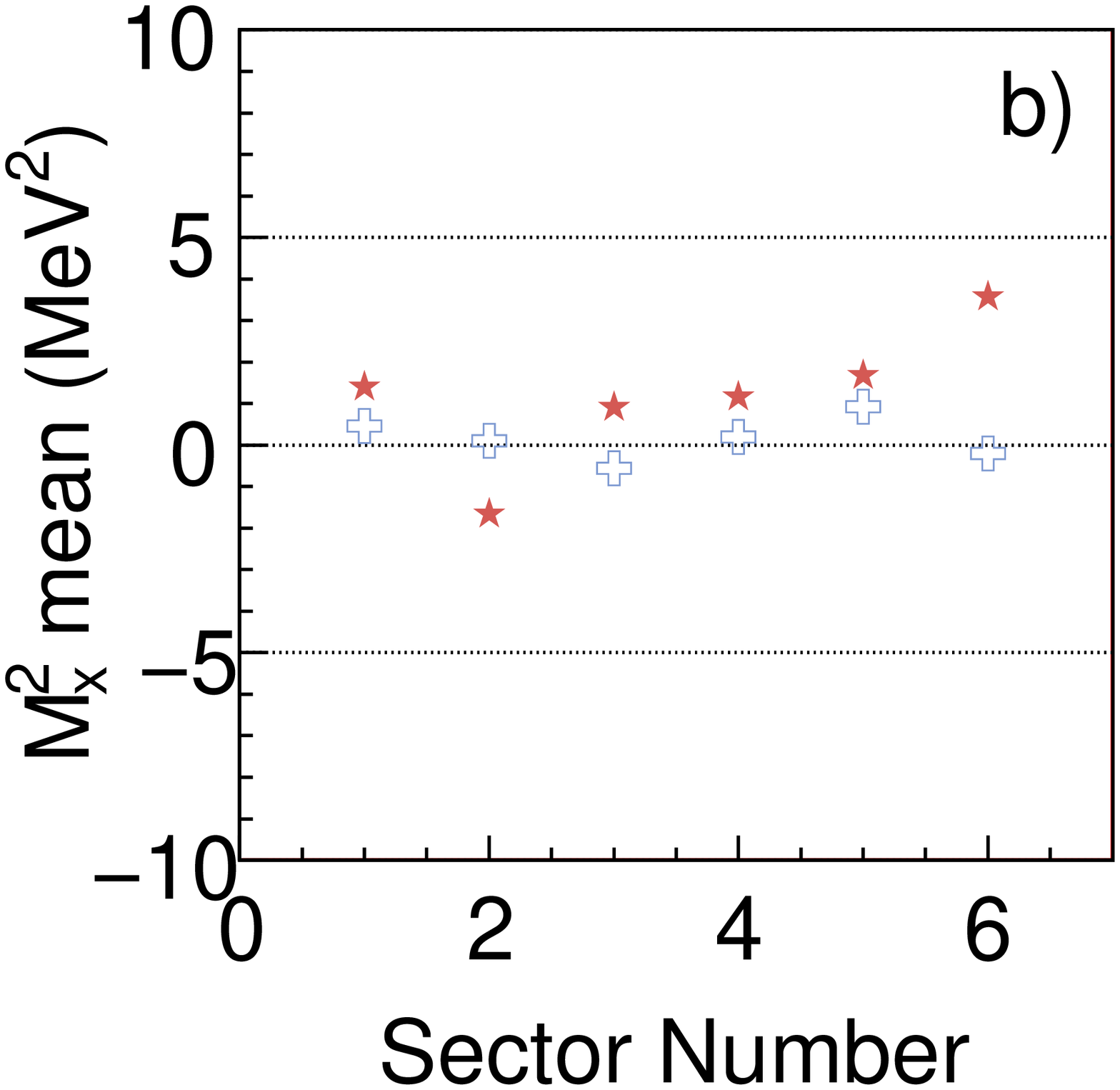}}
{\includegraphics{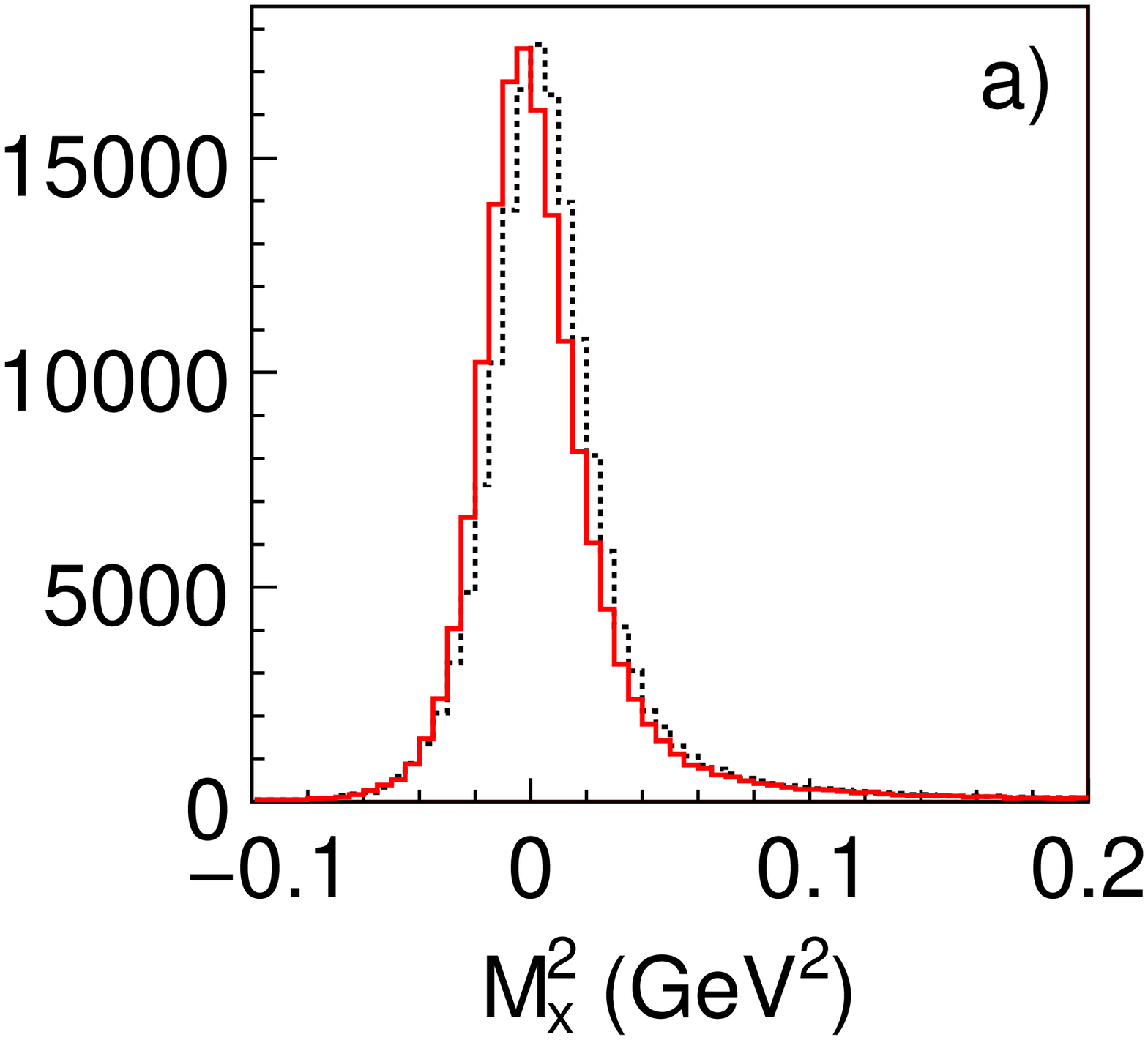}}
\caption{(Color) a) The missing mass squared of
$ep\to epX$ for the BH events before (the dashed line
histogram) and after (the solid line histogram) momentum corrections are
applied. b) The dependence of the Gaussian means of the missing mass squared
distributions on the CLAS sector before (asterisks) and
after (crosses) the momentum corrections.}
\label{fig:mmpi0}
\end{figure}

Both methods give comparable results. For the final analysis,
corrections obtained by the second method are used. In
Fig.~\ref{fig:mmpi0} a) the missing mass squared distribution for BH
events is shown before (dashed line histogram) and after (solid line
histogram) corrections are applied. After corrections are applied
the position of the missing mass peak has moved to the correct 
value and the width of the peak has improved. In
Fig.~\ref{fig:mmpi0} b) the improvement on the mean values of the missing mass ($M_x^2$)
distributions of BH events as a function of CLAS sector are
shown. The variation of the centroids is less than $0.001$ GeV$^2$
after momentum corrections. 

\section{DVCS Analysis}\label{DVCS_ANALYSES_CHAPTER}


\indent

\begin{figure*}[ht!]
\vspace{90mm} 
{\includegraphics{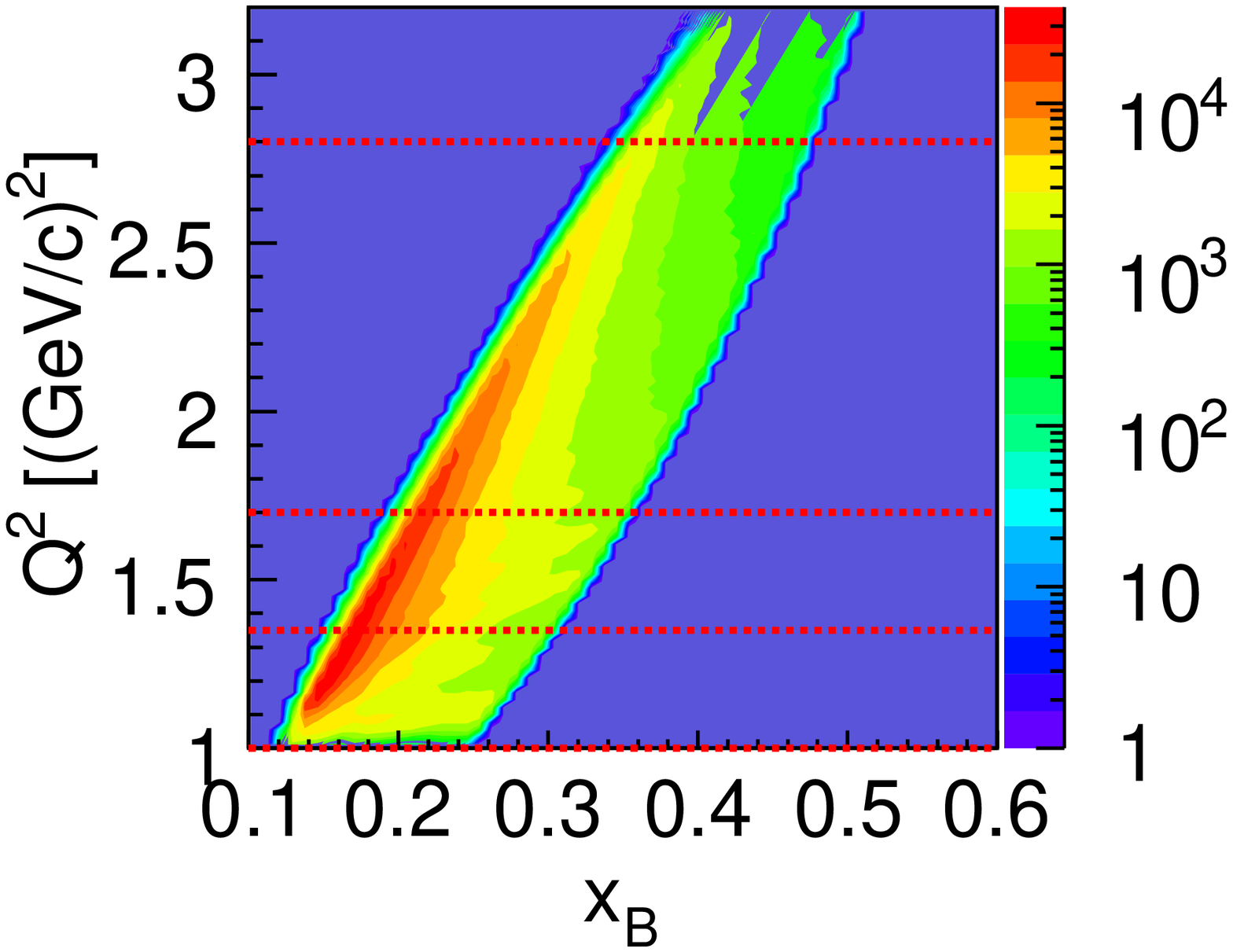}}
{\includegraphics{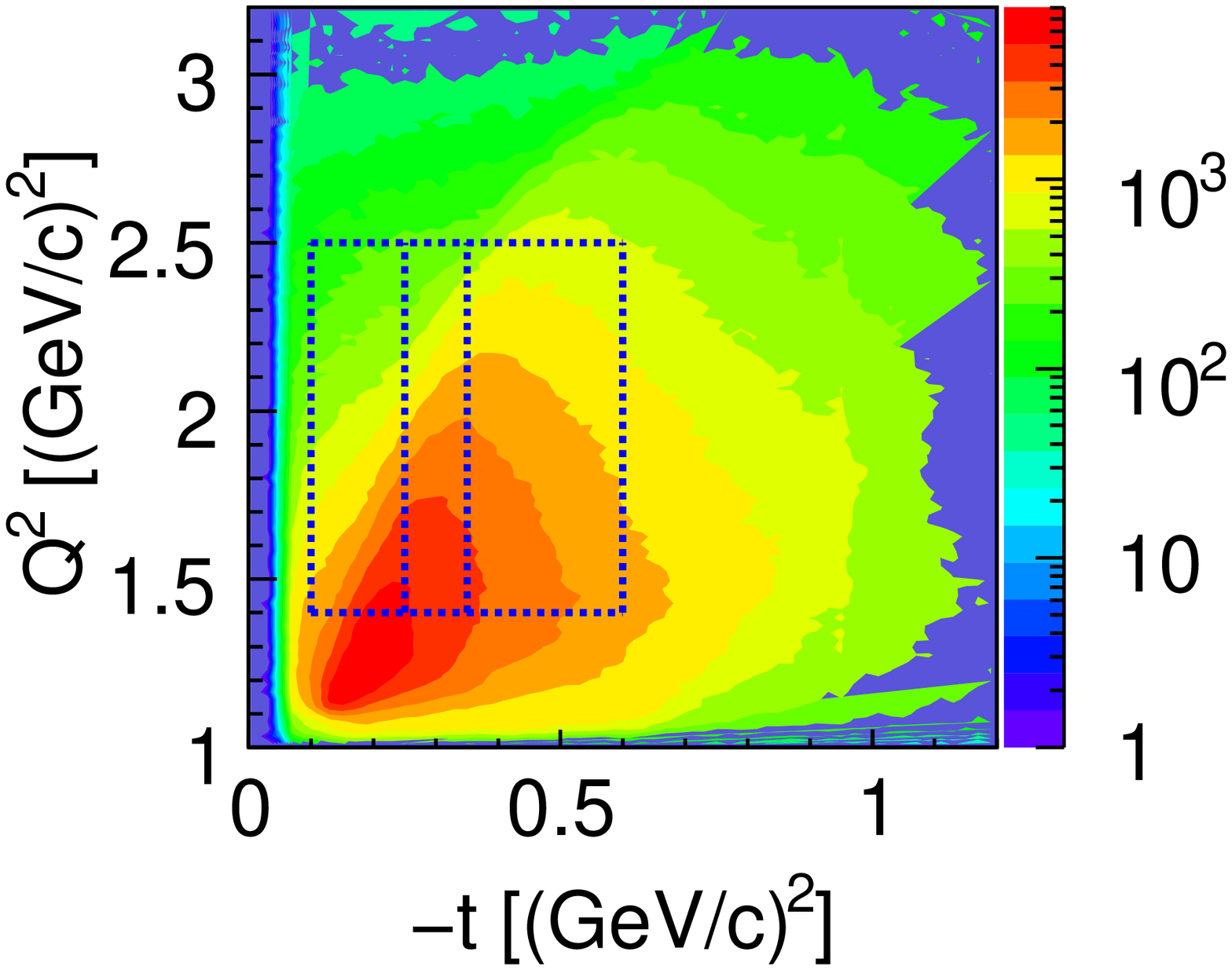}}
\caption{\label{fig:KINEMATICS_Q2_T_XB} (Color) The kinematical coverage of
the data set. (Left) The $Q^2$ vs. $x_{B}$ distribution is shown
with our $Q^2$ bin divisions and (right) the $Q^2$ vs. t distribution is shown.
The boxes indicate the kinematical bins used for this analysis (see
Tables~\ref{tab:ASYM_ALL_RESULTS_Q2} and ~\ref{tab:ASYM_ALL_RESULTS_t}).} 
\end{figure*}

The beam spin asymmetry in DVCS is studied in the kinematical region of  $W>2$ GeV and
$1.4$ (GeV/c)$^2$ $< Q^2 < 2.8 $ (GeV/c)$^2$
 using the reaction \EpX  where only one electron and one
proton are detected in CLAS. The kinematical coverage is shown in  Fig.~\ref{fig:KINEMATICS_Q2_T_XB}.

The azimuthal angular modulations 
of the beam spin asymmetry ($A_{LU}$) were extracted for three bins in
$Q^2$, shown with horizontal lines in the left graph of Fig.~\ref{fig:KINEMATICS_Q2_T_XB}, for
the transferred momentum range from $0.1$ (GeV/c)$^2$ to $0.4$
(GeV/c)$^2$. The $t$ dependence of the angular modulations have been studied
in three bins of $t$, shown by the vertical lines in the right graph, for the $Q^2$ range from $1.4$
(GeV/c)$^2$ to $2.5$ (GeV/c)$^2$.     

\begin{figure}[ht!]
\vspace{47mm} 
{\includegraphics{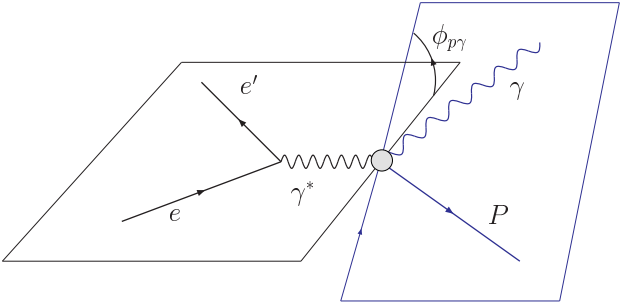}}
\caption{\label{fig:phidef} (Color) The kinematics of electroproduction in
the target rest frame. The azimuthal angle $\phi_{p\gamma}$ is the angle
between the proton-photon production plane and the electron scattering plane.
The azimuthal angle defined in Ref.~\cite{bel0112} and presented in
Eq.(\ref{eq:sfm}) is $\phi=\pi-\phi_{p\gamma}$.} 
\end{figure}

For each $Q^2$ and $t$ bin, the data were divided into
11 bins of $\phi_{p\gamma}$, the azimuthal angle of the proton-photon
production plane to the electron scattering plane (see
Fig.~\ref{fig:phidef}). Note that in 
our notation $\phi$ is defined as $\phi=\pi-\phi_{p\gamma}$. In each kinematical bin, the 
number of \pEpg~ events for positive ($N_\gamma^+$) and negative
($N_\gamma^-$) helicity states of the beam electrons and for the
helicity sum ($N_\gamma$) have been extracted using a fit to the
line shape of the missing mass squared distributions. The beam spin
asymmetry was calculated as:
\begin{equation}
A_{LU} = \frac{1}{P_e} \cdot \frac{N^+_{\gamma} - N^-_{\gamma}
}{N_{\gamma} } ,
\label{eq:CACL_ASYM}
\end{equation}
where $P_e$ is the average polarization of the beam. 
The helicity-related charge asymmetry has been measured using the inclusive
electron yield. It was found to be $0.7\%$ and is
included in the estimation of the systematic uncertainties.
 
 After omitting terms suppressed by an order of $-t/Q^2$ or higher in Eqs.(\ref{Par-BH}), (\ref{AmplitudesSquared}), (\ref{InterferenceTerm}), and
(\ref{eq:sfm}), $A_{LU}$ can be written as:
\begin{eqnarray}
\label{eq:ASYMSINp}
A_{LU} &=& {\alpha^\prime \sin \phi \over {1+\gamma \cos \phi}},
\end{eqnarray}
where the coefficients $\alpha^\prime$ and $\gamma$ are:
\begin{eqnarray}
\label{note}
\alpha^\prime &=& \frac{\Bx }{y}(1 + \epsilon^2)^2 \frac{s_1^{\cal I}} {c^{\rm BH}_0 } \\ \,
\gamma &=& \frac{y c^{\rm BH}_1  +\Bx (1 + \epsilon^2)^2c_1^{\cal I}}{y c^{\rm BH}_0} .
\end{eqnarray}

In our analysis, the $\sin{\phi_{p\gamma}}$ moment 
of $A_{LU}$, which is a linear combination of GPDs (see
Eq.\ref{eq:sfm}), has been extracted in each 
$Q^2$ and $t$ bin using fits to the $\phi_{p\gamma}$ dependence of $A_{LU}$ 
with the function: 
\begin{eqnarray}
\label{eq:ASYMSIN}
A_{LU} &=& \alpha \sin \phi_{p\gamma} + \beta \sin 2\phi_{p\gamma} .
\end{eqnarray}
In this representation all the higher-order terms in the azimuthal angle
have been combined in the parameter $\beta$. 
Note that power-suppressed contributions could modify Eqs. (\ref{eq:intxt14}) and (\ref{eq:sfm}) \cite{DM_Private}.

\subsection{The Missing Mass Technique}
\label{mmtech}
\indent

The main challenge in the DVCS analysis using the reaction \EpX
is the separation of the single photon events from the more
than one photon events, mostly $\pi^0$.
In the kinematics of deep inelastic scattering, the CLAS
resolution of the missing mass squared is not sufficient to separate
 these two final states cleanly. For the separation of 
$ep\gamma$ and $ep\gamma\gamma$ events, a fit to the line
shape of the missing mass squared distribution is used.
This technique for the extraction of the DVCS information in
the reaction \EpX~ has already been used in the DVCS analysis of the CLAS
data at $4.2$ GeV beam energy \cite{Clas01a}. The high statistics of this
data set and the comprehensive simulation, enable detailed
studies of the systematic uncertainties associated with the missing mass
technique.
 
\begin{figure*}[ht]
\vspace{90mm} 
{\includegraphics{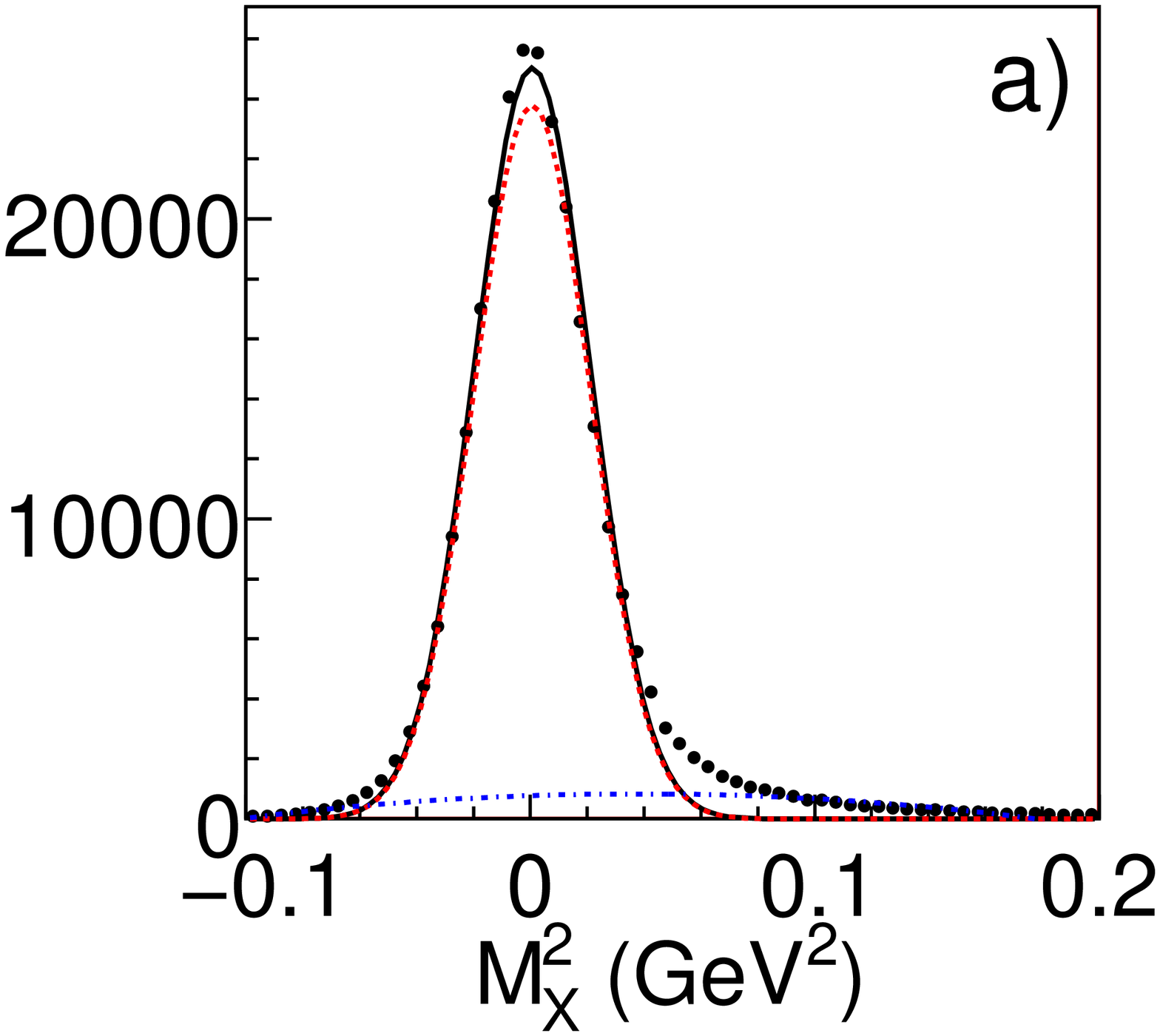}}
{\includegraphics{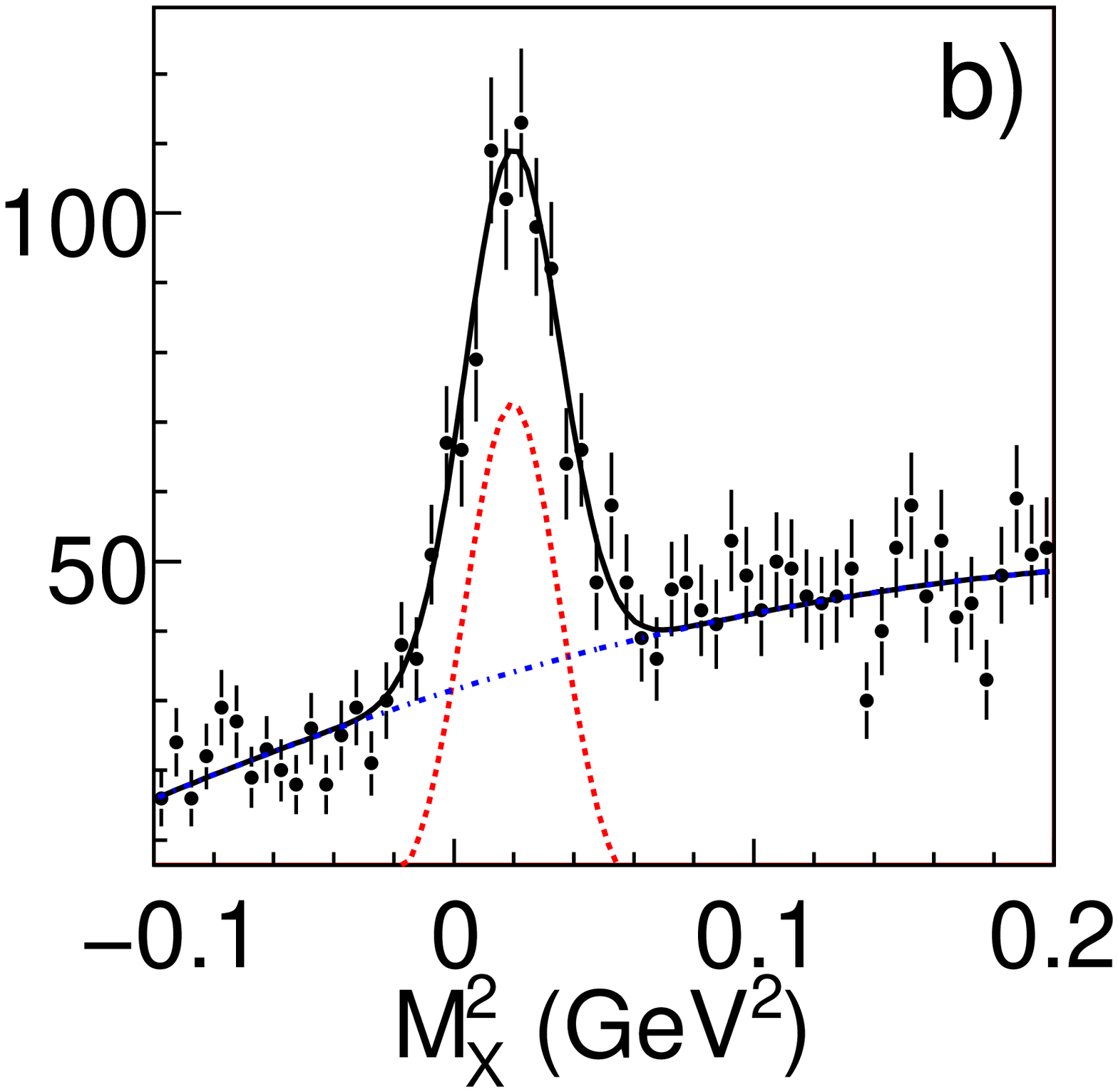}}
\caption{(Color) Fits to identified $ep\gamma$ a) and
$ep\pi^0$ b) final states to determine the mean values and standard
  deviations of the Gaussian functions  
$G_{\gamma}$ and $G_{\pi^o}$. The solid line is the fit function, the
dashed line is the fitted Gaussian 
function, and the dashed-dotted line is the polynomial fit to the
radiative tail and the background.} 
\label{fig:gpimm}
\end{figure*}

The main contributions to the missing mass squared distribution in the
range $-0.1$ GeV/c$^2$$<M_x^2<0.2$ GeV/c$^2$ come
from three processes: 1) single photon production; 2)
$\pi^0$ production; and 3) the radiative processes associated with photon and $\pi^0$ production, such as the
$ep\gamma_R\gamma$ or $ep\gamma_R\pi^0$ final states. Here $\gamma_R$
is the photon radiated by the incoming or outgoing electron. 
There are also background events associated with particle
misidentification and multi-pion production. In the fit, it is assumed 
that the missing mass squared distributions corresponding to 
single photon and single $\pi^0$ production have a
Gaussian shape defined by the detector resolution. 
The radiative tails associated with $\gamma$ and $\pi^0$ production,
and the background from particle misidentification and multi-pion
production, is fitted by a polynomial function.
A sum of two Gaussians and a polynomial function is used in the final fit,
\begin{eqnarray}
F = N_{\gamma} \cdot G_{\gamma} + N_{\pi^0} \cdot G_{\pi^0} + P_3\cdot
Pol_3 ,
\label{eq:g2p}
\end{eqnarray}
where the fit parameters $N_{\gamma}$ and $N_{\pi^{0}}$ are the number of
 single photon and single pion events, 
respectively. The fit parameter
$P_{3}$ is the 
relative magnitude of the background. The shape of the background,
$Pol_3$, has been determined by a fit to the tails of the missing mass
squared distribution. The mean values and the standard deviations of the
Gaussian functions 
$G_{\gamma}$ and $G_{\pi^{0}}$ are determined by the fits to the
missing mass squared distributions of the BH, $ep
\rightarrow e'p\gamma_R$, and the single pion, $ep \rightarrow
e'p{\pi^{0}}$, events. 
The identification of the BH and $\pi^0$ events is conducted
the same way as was described above. For the extraction of the Gaussian
parameters, the function $f = A \cdot
G_{\gamma (\pi^o)}(M_x,\sigma_x) + Pol_3$ is used to fit the missing
mass squared distributions of identified BH and $\pi^0$ final states individually.
Here, the fit parameter $A$ is the amplitude of the Gaussian function.
The parameters $M_x$ and $\sigma_x$ are the mean value and the standard
deviation of the Gaussian function. The function $Pol_3$, as in
Eq.(\ref{eq:g2p}), is used to describe the radiative tail and the background under the
missing mass peak. An example of a fit to the missing mass squared
distribution of $ep\to epX$ for the BH events in the kinematical region
$1.0$ (GeV/c)$^2$$<Q^2<1.35$ (GeV/c)$^2$ and $0.1$ (GeV/c)$^2$$<-t<0.4$ (GeV/c)$^2$ is shown in
Fig.\ref{fig:gpimm}a. A fit to
the $M_x^2$ distribution of the single pion events for the 
same kinematical bin is shown in Fig. \ref{fig:gpimm}b.
The parameters $M_x$ and $\sigma_x$ have been determined
for each $Q^2$ and $t$ bin.

\begin{figure*}[ht!]
\vspace{145mm} 
{\includegraphics{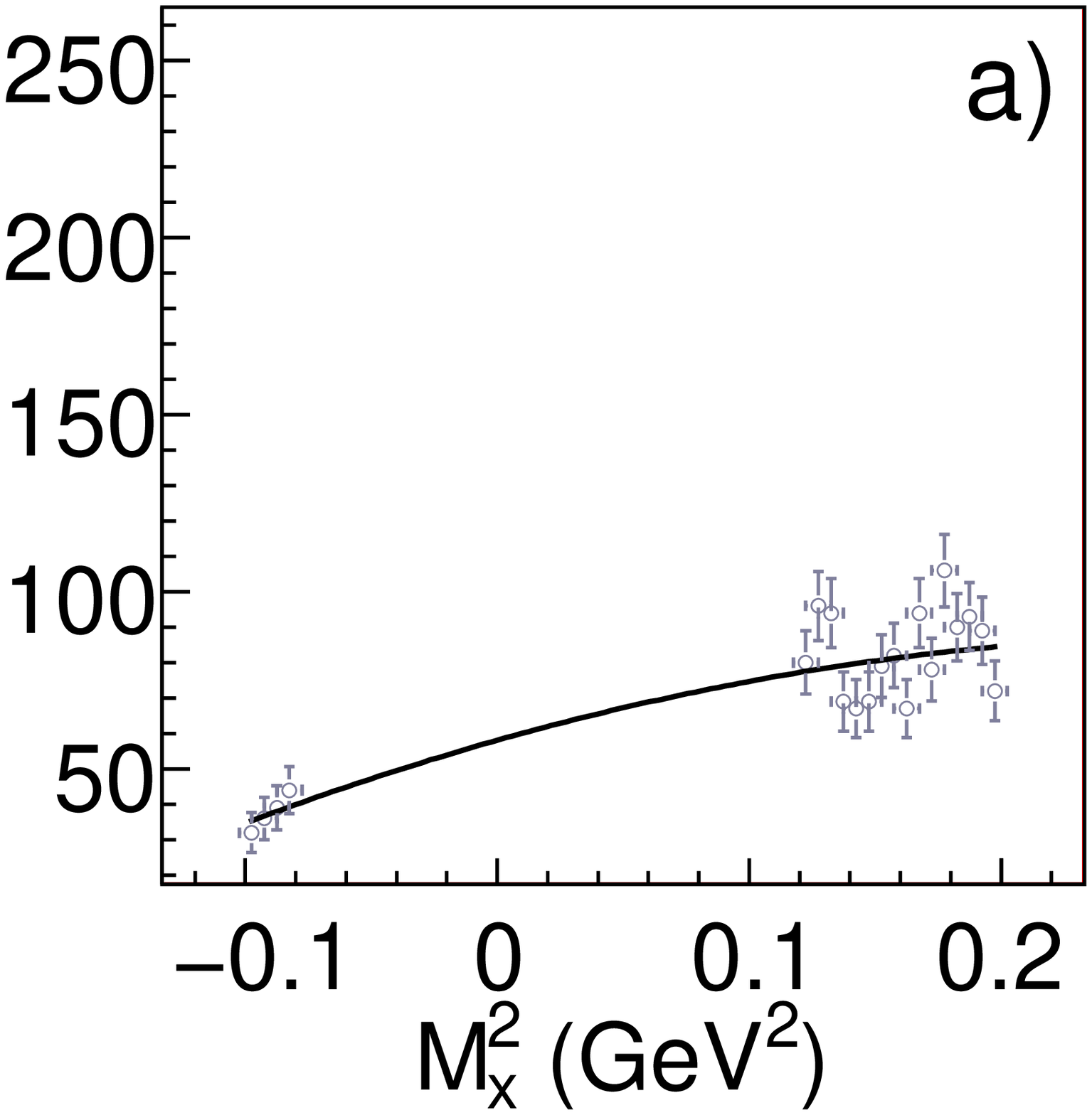}}
{\includegraphics{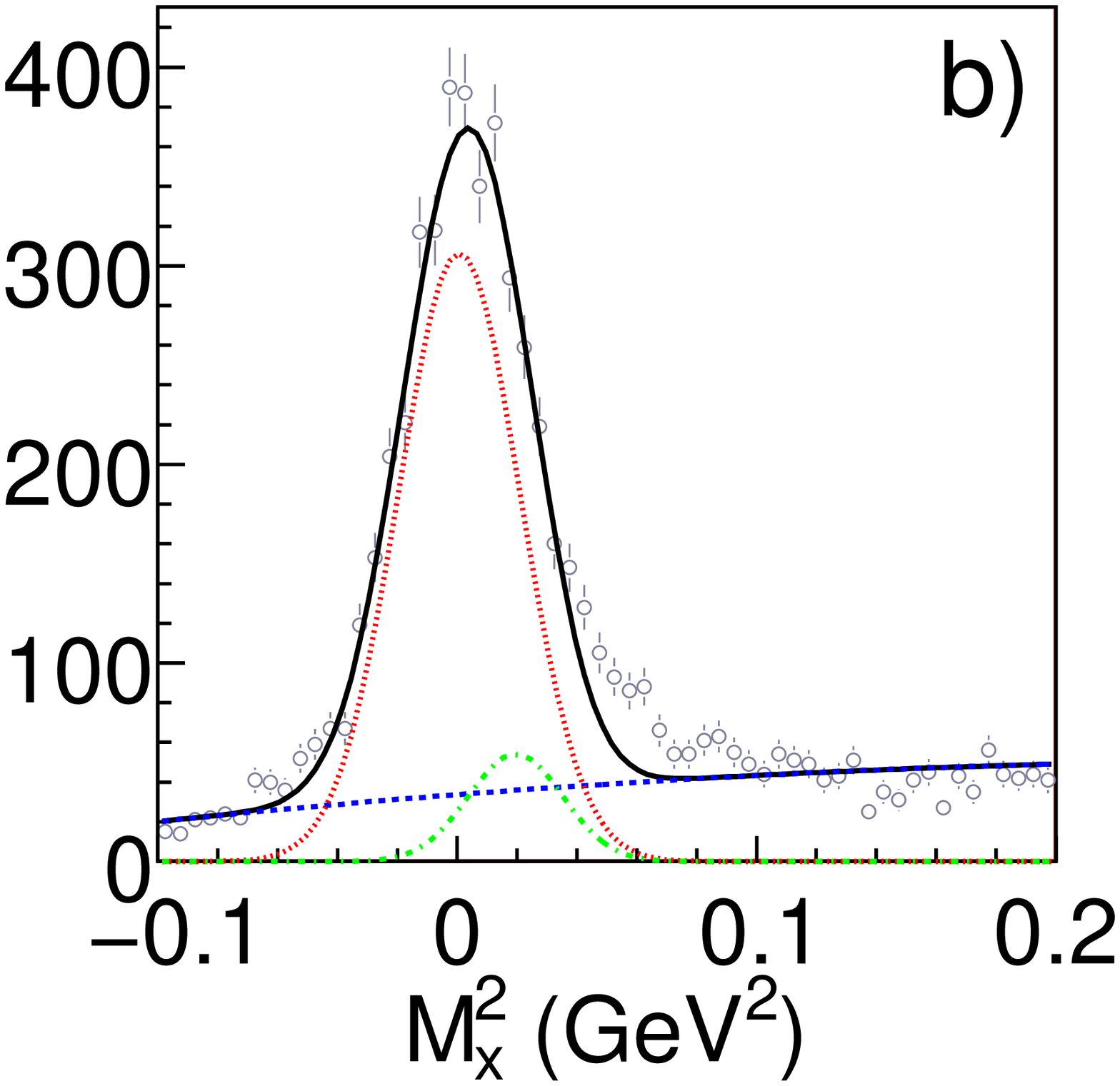}}
{\includegraphics{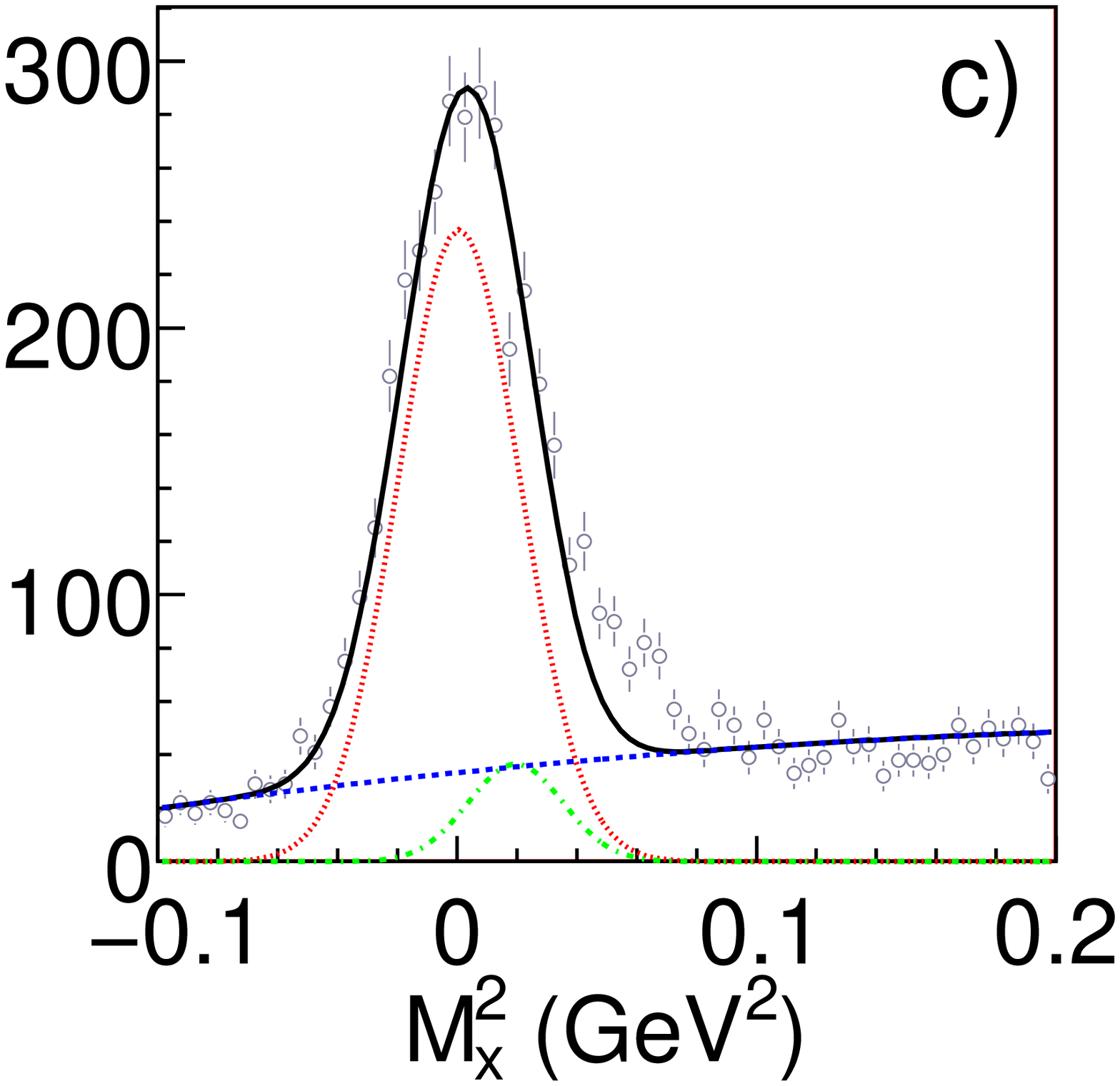}}
{\includegraphics{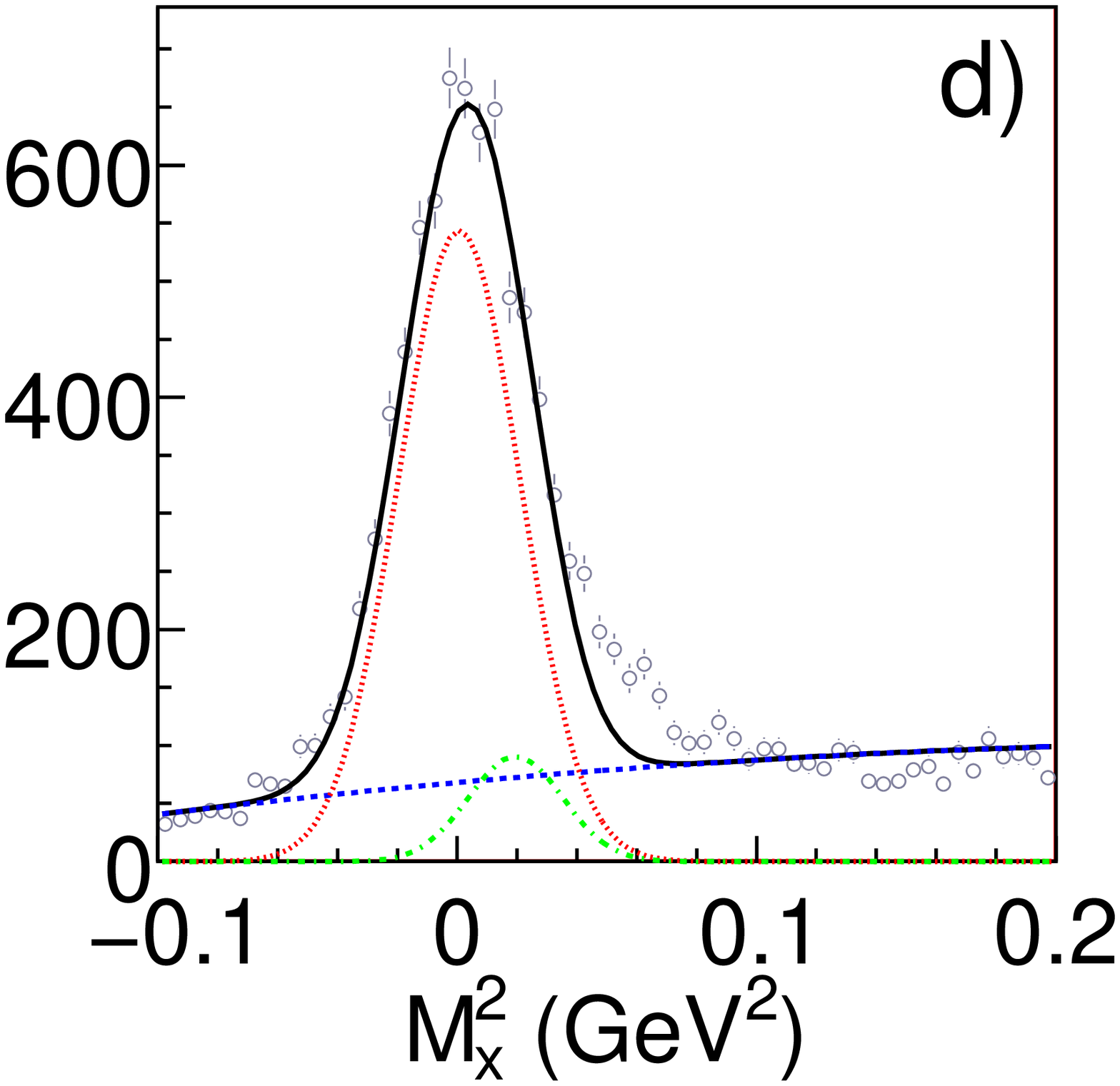}}
\caption{(Color) Example of fits obtained using Eq.~(\ref{eq:g2p}) for one 
$\phi_{p\gamma}$ bin. a)  Fit to the background using the end points of
the $M_x^2$ distribution, b), c), and d) are the fits to the $M_x^2$ distributions
corresponding to positive, negative, and summed beam helicity states, respectively. 
In b), c), and d) the fit parameters are the number of single photon
and $\pi^0$ events.} 
\label{fig:mmfit}
\end{figure*}

The missing mass squared distributions for each helicity state and for
the helicity sum in each kinematical bin are fitted using the function in
Eq.(\ref{eq:g2p}). 
An example of a fit for a typical bin, 
$\phi_{p\gamma}$ from $35^o$ to $70^o$,  $1.0$ (GeV/c)$^2$$<Q^2<1.35$ (GeV/c)$^2$, and
$0.1$ (GeV/c)$^2$$<-t<0.4$ (GeV/c)$^2$ is shown in 
Fig.~\ref{fig:mmfit}. First, a fit with a third order polynomial is
performed to the points outside of the peak region, $-0.1$ (GeV/c)$^2$$<M_x^2<-0.08$ (GeV/c)$^2$
 and
$0.12$ (GeV/c)$^2$$<M_x^2<0.2$ (GeV/c)$^2$, to determine the shape of the polynomial $Pol_3$, 
see Fig.~\ref{fig:mmfit}a. Then,  
using the corresponding mean values and standard 
deviations of the Gaussian functions $G_\gamma$ and $G_{\pi^0}$,
fits to the $M_x^2$ distributions for the positive 
(Fig.~\ref{fig:mmfit}b) and negative 
(Fig.~\ref{fig:mmfit}c) beam helicity states, and for the
helicity sum (Fig.~\ref{fig:mmfit}d) were performed. In
Fig.~\ref{fig:mmfit}b, Fig.~\ref{fig:mmfit}c, and Fig.~\ref{fig:mmfit}d the
solid lines are the fit functions as defined in Eq.(\ref{eq:g2p}), the
dashed lines are the Gaussian functions for the single photon events, the dashed-dotted
lines correspond to the Gaussian functions for the single $\pi^0$ events, and the
dashed-dot-dot-dot lines correspond to the polynomial background.

%
%
%
Fits to the  $M_x^2$ distributions
have been performed
in 11 $\phi_{p\gamma}$ bins for each $Q^2$ and $t$ bin.
The extracted number of photon events are used to calculate $A_{LU}$
for each kinematical point. 


\section{Studies of the Analysis Method}

\indent

The mean values of the missing mass squared distributions for
the single photon and $\pi^0$ final states 
are separated by approximately one-$\sigma$ of the fitted Gaussian distributions. This implies that any small systematic 
uncertainties in the energy calibration will directly affect the fit
results. 
The accuracy of the determination of $N_\gamma$ has been 
studied using real and simulated data. In both
cases, single photon and $\pi^0$ events are mixed together and the
missing mass squared distributions of the mixed samples are fitted
using the function in Eq.(\ref{eq:g2p}) to reconstruct the number of
photon and pion events. The difference between the initial and
the reconstructed number of photon events is taken as a measure of
the uncertainty. Using the fit results to the $M_x^2$
distributions for the DVCS analysis as a guide, the mixed samples
with different statistics and the relative ratios of the photon and pion
events have been studied.

\subsection{Studies With Experimental Data}

\indent

The main advantages of using measured data are the correct representation of
the background shape and the missing mass squared resolutions.
First, separate sets of identified
$ep\gamma$ and $ep\pi^0$ final states have been created. The method of
identification of the photon and pion final states has been explained
before. The initial number of events in each final state for the
mixing in a given kinematical bin is
determined by using a single Gaussian and a polynomial fit to the $M_x^2$
distributions. Then the two samples were mixed and 
the number of events in each final state in the mixed sample was then
reconstructed by a fit to the line shape of the missing mass squared
distribution by using the function in Eq.(\ref{eq:g2p}).

\begin{figure*}[ht!]
\vspace{80mm} 
{\includegraphics{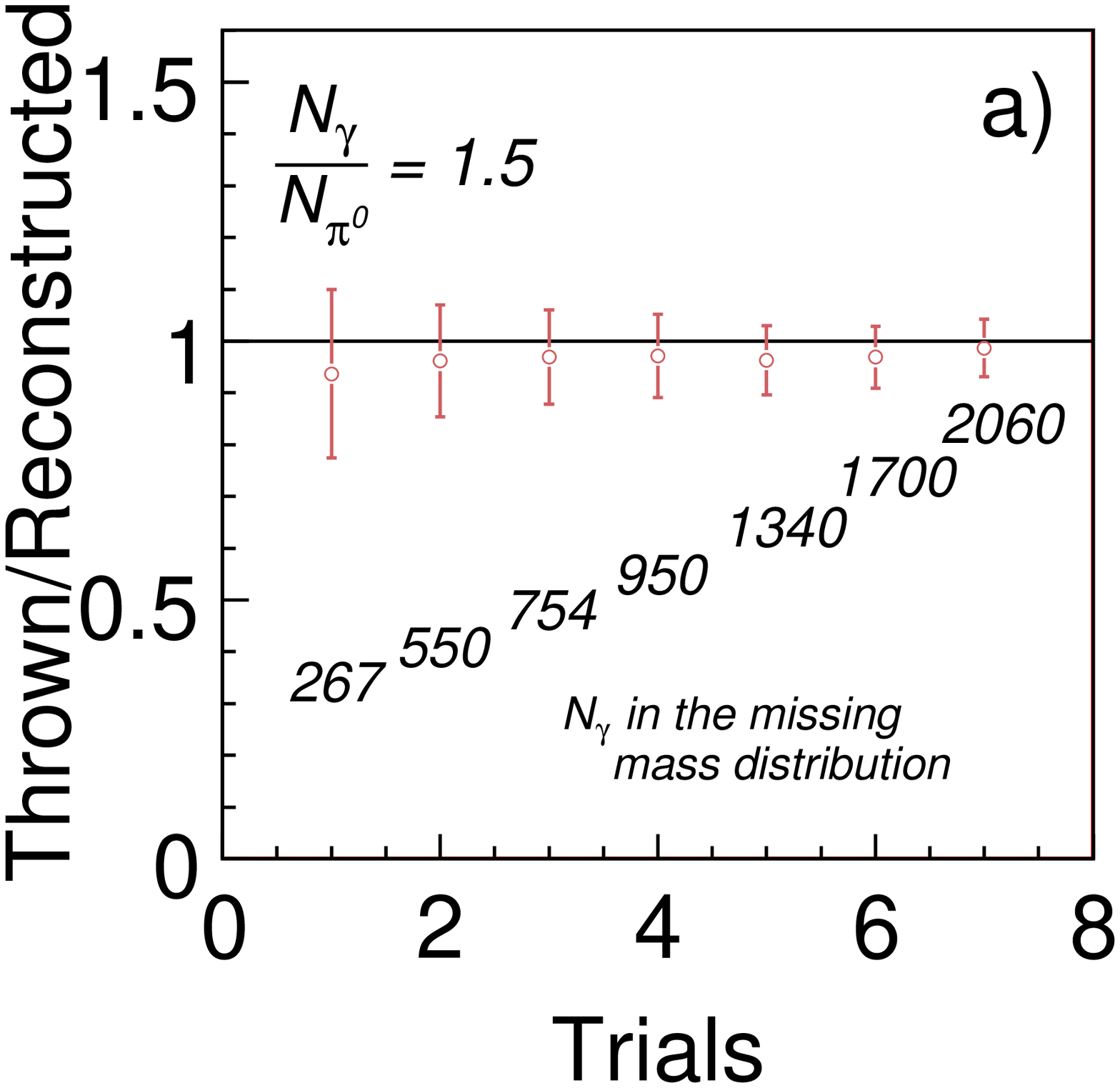}}
{\includegraphics{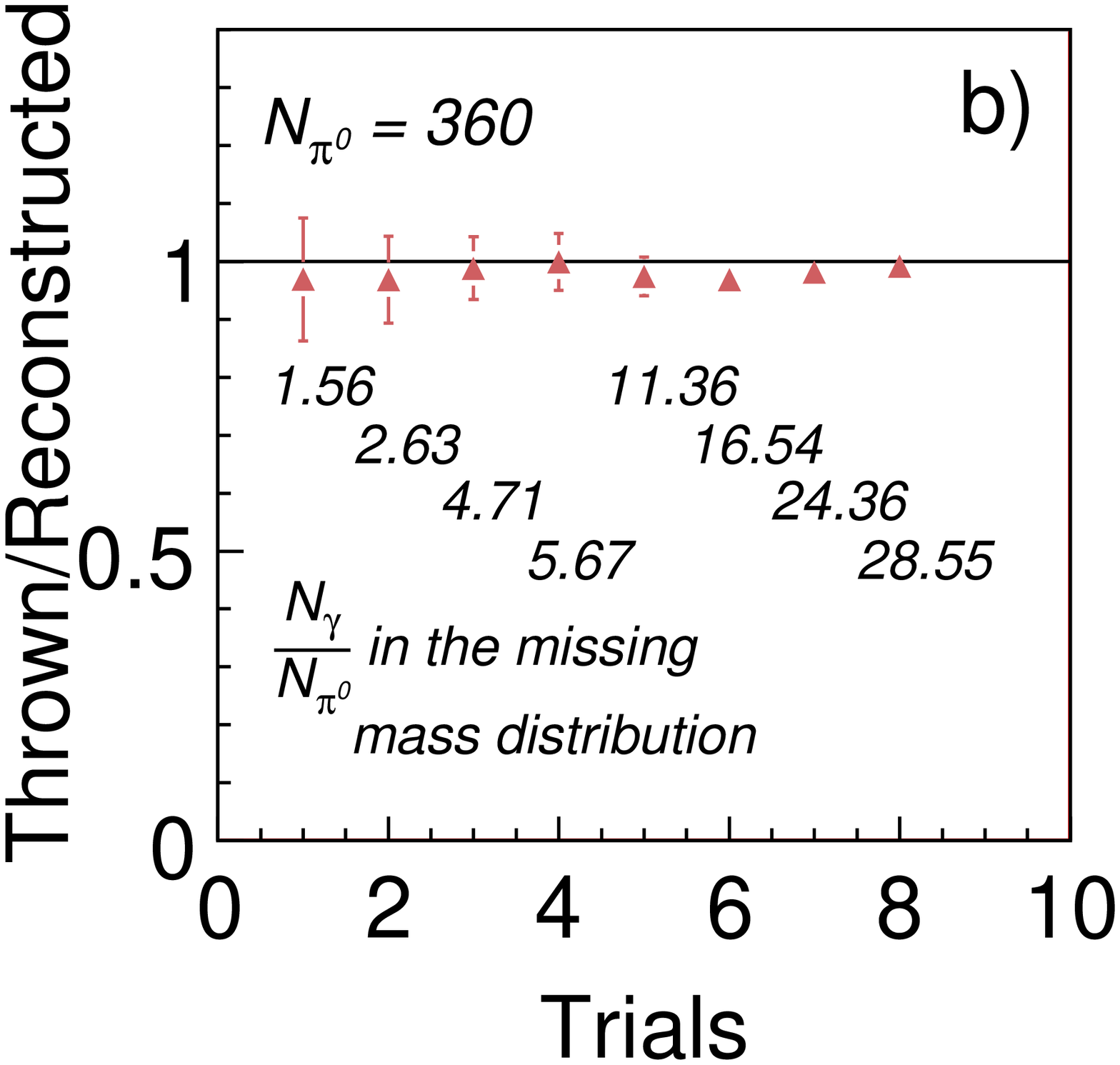}}
\caption{\label{fig:rat1} (Color) Ratios of initial to reconstructed single
  photon events a) for a constant 
  $N_{\gamma}/N_{\pi^0}$ ratio and b) for a constant number of events in
  the mixed distribution. In a) the
numbers under each point on the plot correspond to the initial number
of $ep\gamma$ events. In b) the numbers under each
point show the ratios of the $ep\gamma$ to $ep\pi^0$ events in the
mixed sample.} 
\end{figure*}

Two cases of
event mixing are considered. In the first case, the mixing ratio of the 
$ep\gamma$ and $ep\pi^0$ events is kept constant at $1.5$, while
the total number of events in the mixed distributions was changed.
The red points in Fig.~\ref{fig:rat1}a show the ratio of the initial
number of photon events to the 
number of reconstructed events (from a fit to the $M_x^2$
distribution of mixed events). The
numbers under each point on the plot correspond to the initial number
of $ep\gamma$ events. Within statistical uncertainties, the reconstructed and
the initial number of photon events are the same, although there is a
general trend of the reconstructed numbers to be a few percent larger
than the initial number. 

In the second case, the number of $ep\pi^0$ events was kept constant
at $360$, while the number of $ep\gamma$ events increased with each trial.
In Fig.~\ref{fig:rat1}b the ratio of the
initial to the reconstructed number of single photon events is shown
with red points. The numbers under each
point show the ratio of the $ep\gamma$ to $ep\pi^0$ events in the
mixed sample. Again, within statistical uncertainties, the ratios of
the initial to reconstructed number of single photon events are consistent with $1$.

The studies with mixed samples of identified photon and pion final
states using the experimental data show that the number of single
photon events can be reconstructed using the fit to the $M_x^2$
distribution with better than $5$\% accuracy.


\subsection{Studies With Simulated Data}

\indent

Another set of tests has been performed using a realistic event generator \cite{AVAK_DVCS}
for the $ep\rightarrow ep\gamma$ and $ep\rightarrow ep\pi^0$
reactions. The response of the CLAS
detector is simulated using a GEANT-based program. The simulated
data have been processed with the same event reconstruction and physics analysis
algorithms that have been used for the measured data. 
Events from both samples (that passed all analysis cuts) in a given kinematical bin are mixed
together and fits to the line shape of the $M_x^2$ distributions
of the mixed events have been  performed to extract the number of single
photon and
single $\pi^0$ final states. The parameters for $G_\gamma$ and $G_{\pi^0}$
are derived from the fits to the missing mass squared distributions of the  $ep\rightarrow ep\gamma$  and
$ep\rightarrow ep\pi^0$ events before mixing, respectively.  

\begin{figure*}[ht!]
\vspace{80mm} 
{\includegraphics{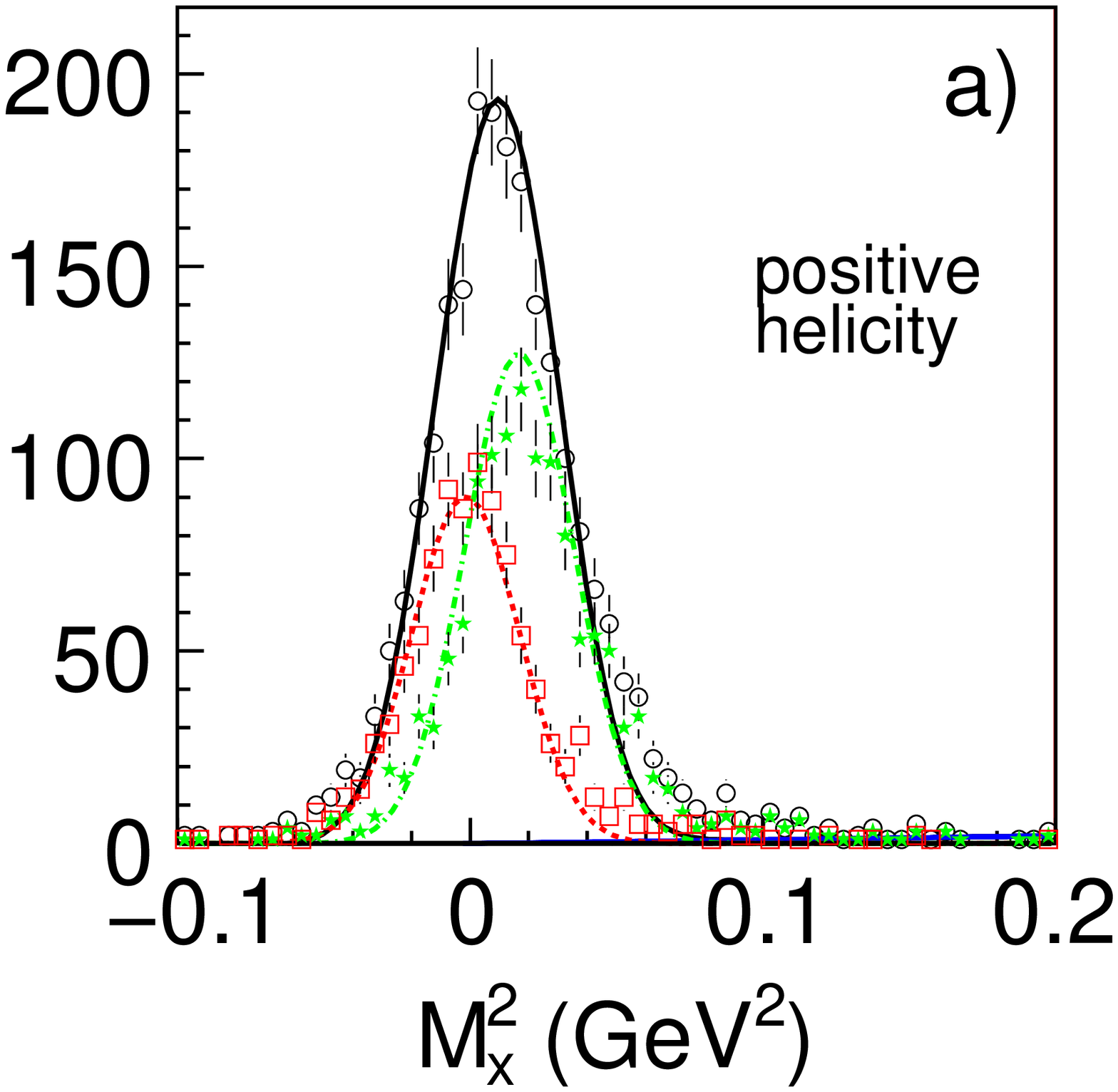}}
{\includegraphics{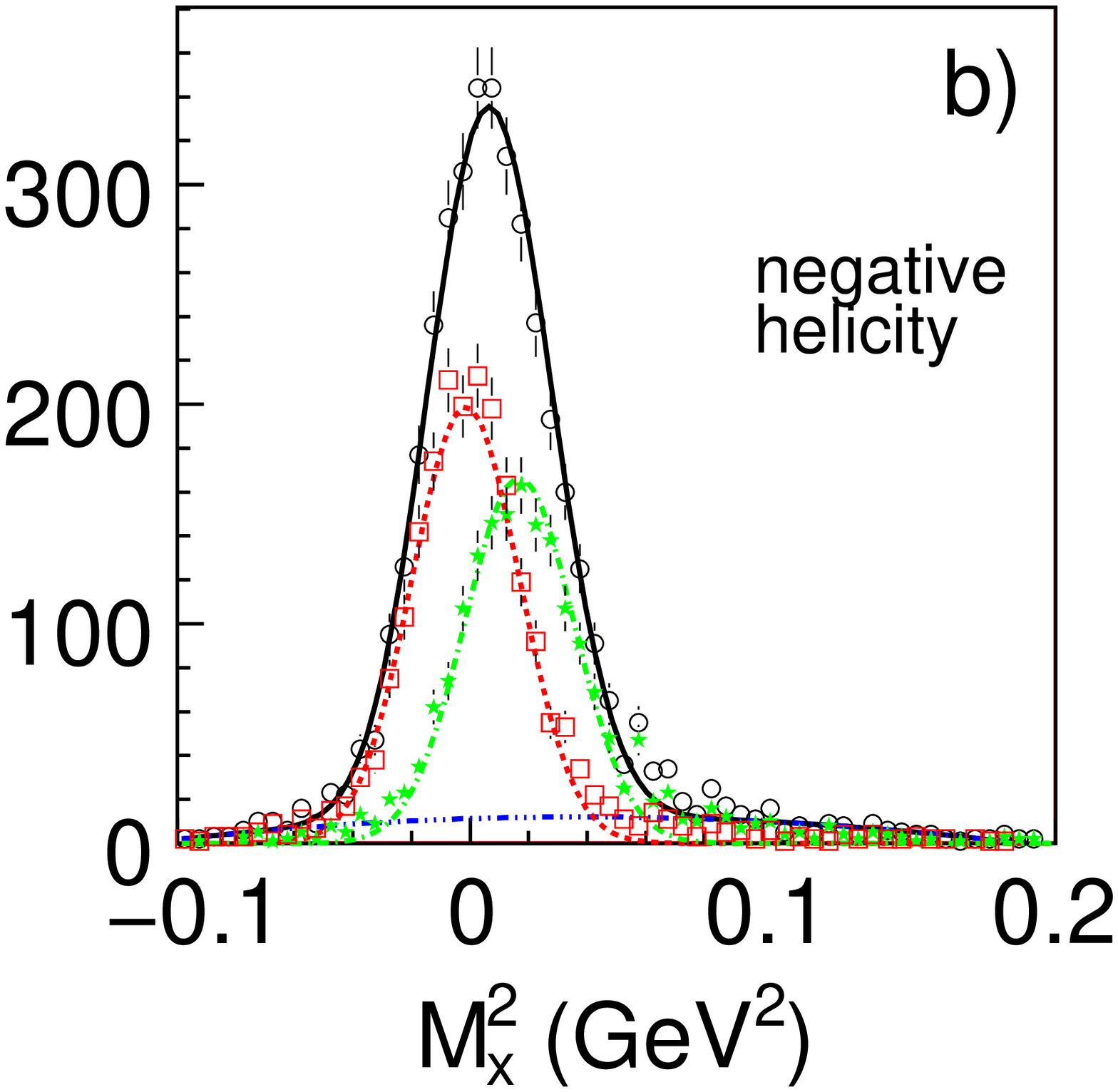}}
\caption{\label{fig:gsim_bin_fits} (Color) Simulated missing mass squared
distributions for positive a) and negative b) helicity for one
$\phi_{p\gamma}$, $Q^2$, and $t$ bin.  
The curves are results of fits using the function in Eq.(\ref{eq:g2p}).
In the simulation radiative effects are not included.}
\end{figure*}

Examples of the fits to the $M_x^2$ distributions for the different
helicity states of the simulated data are shown in
Fig.~\ref{fig:gsim_bin_fits}. 
The open circles with error bars correspond to the mixed event
distributions.
The solid line is the fit using the function in Eq.(\ref{eq:g2p}) to these points. The open squares represent the $M_x^2$
distributions of the initial photon events. The dashed lines are the
resulting $G_\gamma$ from the fit to the mixed distributions.
The asterisks and the dashed-dotted lines are the same for the single pion final
state. The dashed-dot-dot-dotted line (shown only in
Fig.~\ref{fig:gsim_bin_fits}b) corresponds to the fitted polynomial
background. 
\begin{figure}[ht!]
\vspace{100mm}
{\includegraphics{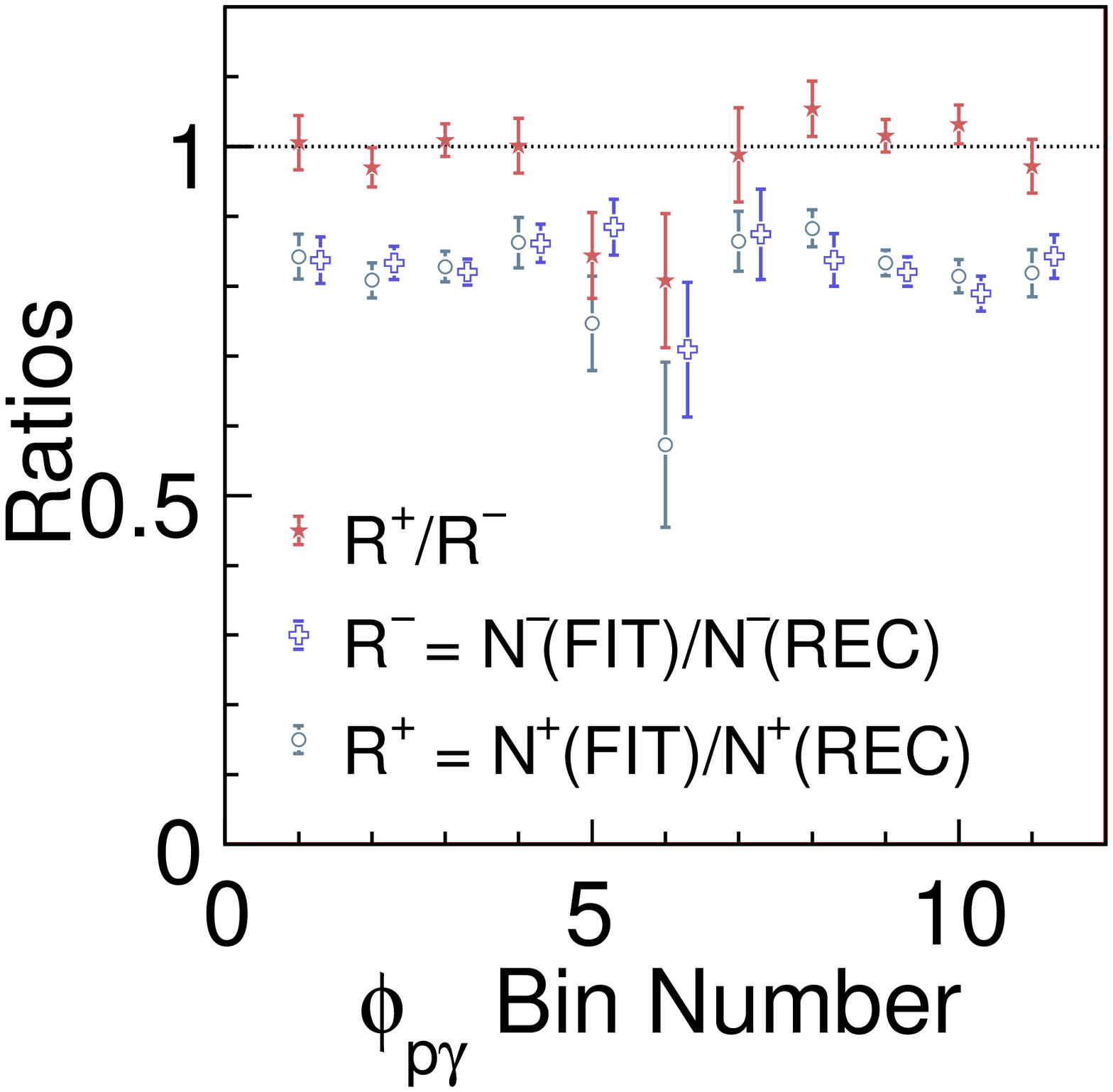}}
\caption{\label{fig:ratios_gp} (Color) The $\phi_{p\gamma}$ dependence of ratios of
the number of single photon events extracted using the fits to the missing mass squared distribution of the mixed events to the number of single photon events that passed the analysis cuts for the positive
 (open circles) and for the negative
  (crosses) beam helicity states. The asterisks represent the ratio of
the two ratios.}  
\end{figure}

\begin{figure}[th!]
\vspace{100mm}
{\includegraphics{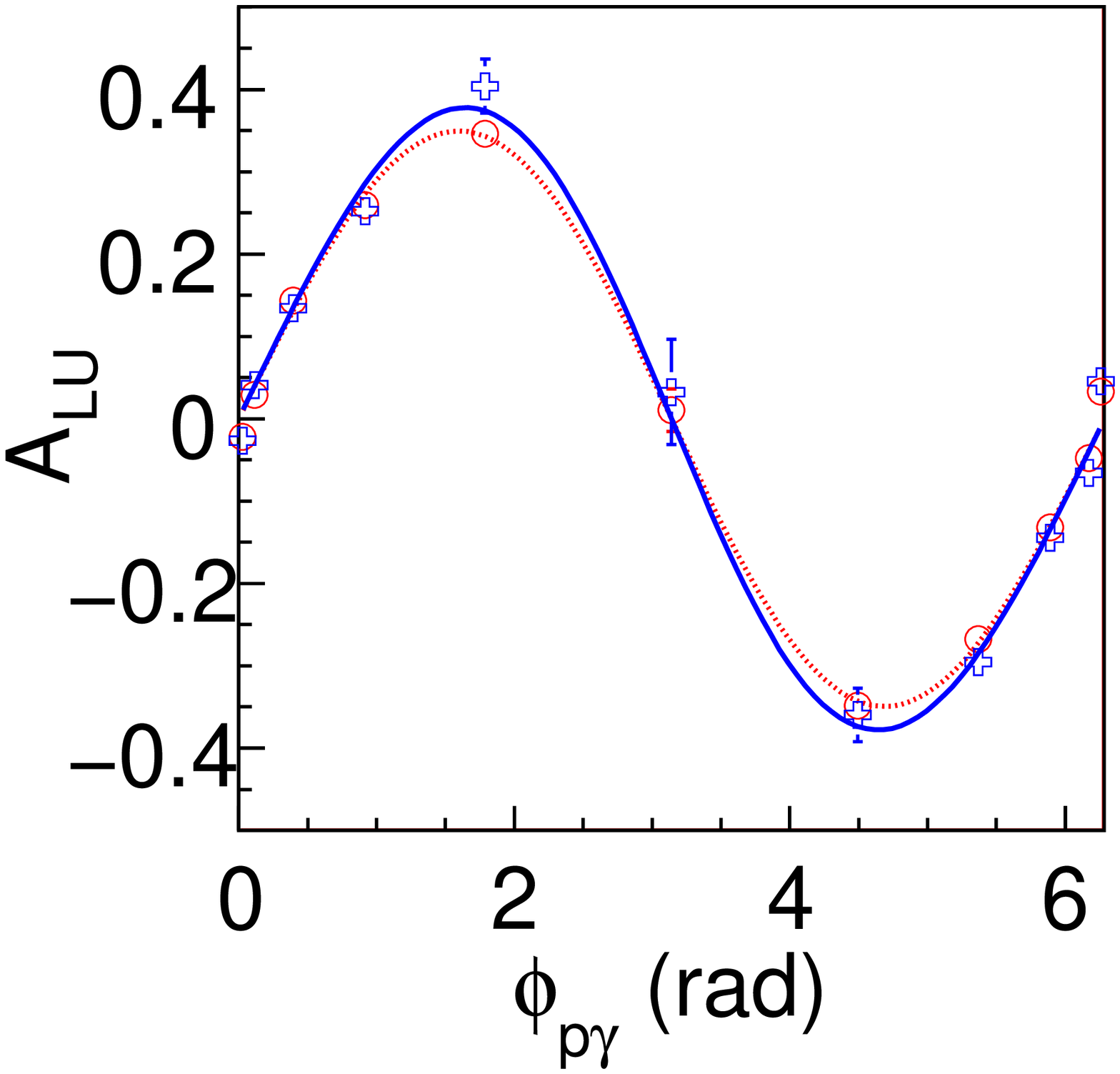}}
\caption{\label{fig:asymetry} (Color) The simulated and extracted
$A_{LU}$. The crosses are the extracted asymmetries and the solid
line is the fit to these points using the function in
Eq.(\ref{eq:ASYMSIN}). The open circles and the dotted line are the same
for the simulated asymmetry.}
\end{figure}

In Fig.~\ref{fig:ratios_gp} the ratio of the number of single photon events,
obtained from fits to the missing mass squared distributions of the mixed event sample, to
the number of reconstructed photon events that passed all analysis cuts are shown for all  
$\phi_{p\gamma}$ bins at $1.0$ (GeV/c)$^2$$<Q^2<1.35$ (GeV/c)$^2$
and $0.1$ (GeV/c)$^2$$<t<0.4$ (GeV/c)$^2$. The crosses and
the circles represent the ratios for different helicities. The
asterisks show the ratio of these ratios. The extracted number of events by the fit is always lower than number of events in the mixed sample by about $20$\%.
This is due to the
fact that there is no background and radiative effects in the
simulated data, while the fit
procedure finds a non-zero contribution for the polynomial function at
 the expense of the photon and 
pion events. The important 
information is indicated by the asterisks, which show that for each
$\phi_{p\gamma}$ bin, these ratios are the same for both helicities.

In Fig.~\ref{fig:asymetry}, the beam spin asymmetry, as defined in
Eq.(\ref{eq:CACL_ASYM}), is presented as a function of
$\phi_{p\gamma}$. The simulated $\phi_{p\gamma}$ dependence of the asymmetry is
shown with circles. The 
crosses represent the asymmetry obtained using the fit procedure. The solid
line is a fit using the function in Eq.(\ref{eq:ASYMSIN}) to the
extracted asymmetry. The
dotted line is a fit to the simulated 
asymmetry. The
$\sin{\phi_{p\gamma}}$ moment for the extracted asymmetry is approximately $3\%$ higher than the simulated
moment. This difference is used as the systematic uncertainty in the
determination of the moments of
the azimuthal asymmetry in DVCS.

\subsection{Dependences on the Parameters of $G_\gamma$ and $G_{\pi^0}$}

\indent

\begin{figure*}[ht!]
\vspace{130mm}
{\includegraphics{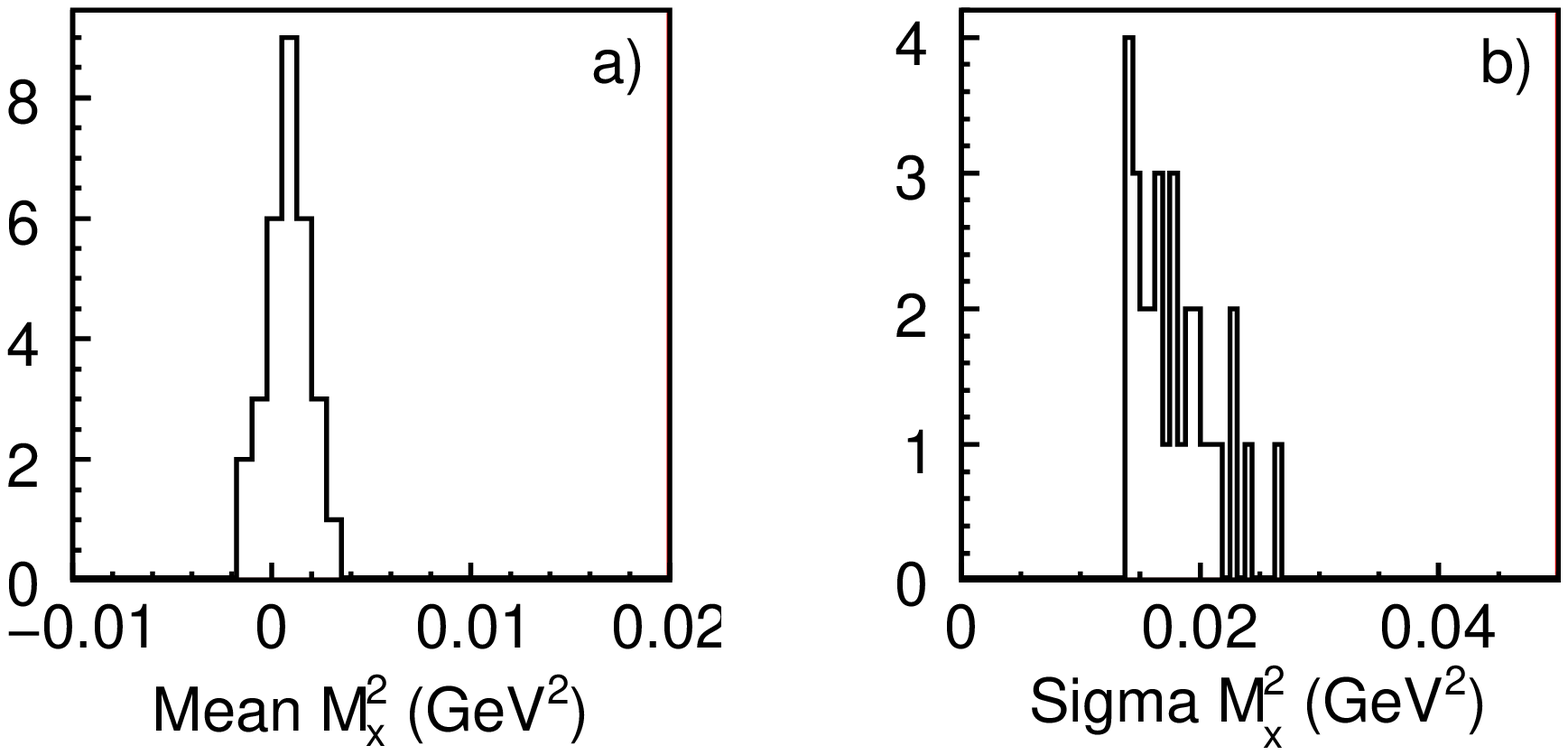}}
{\includegraphics{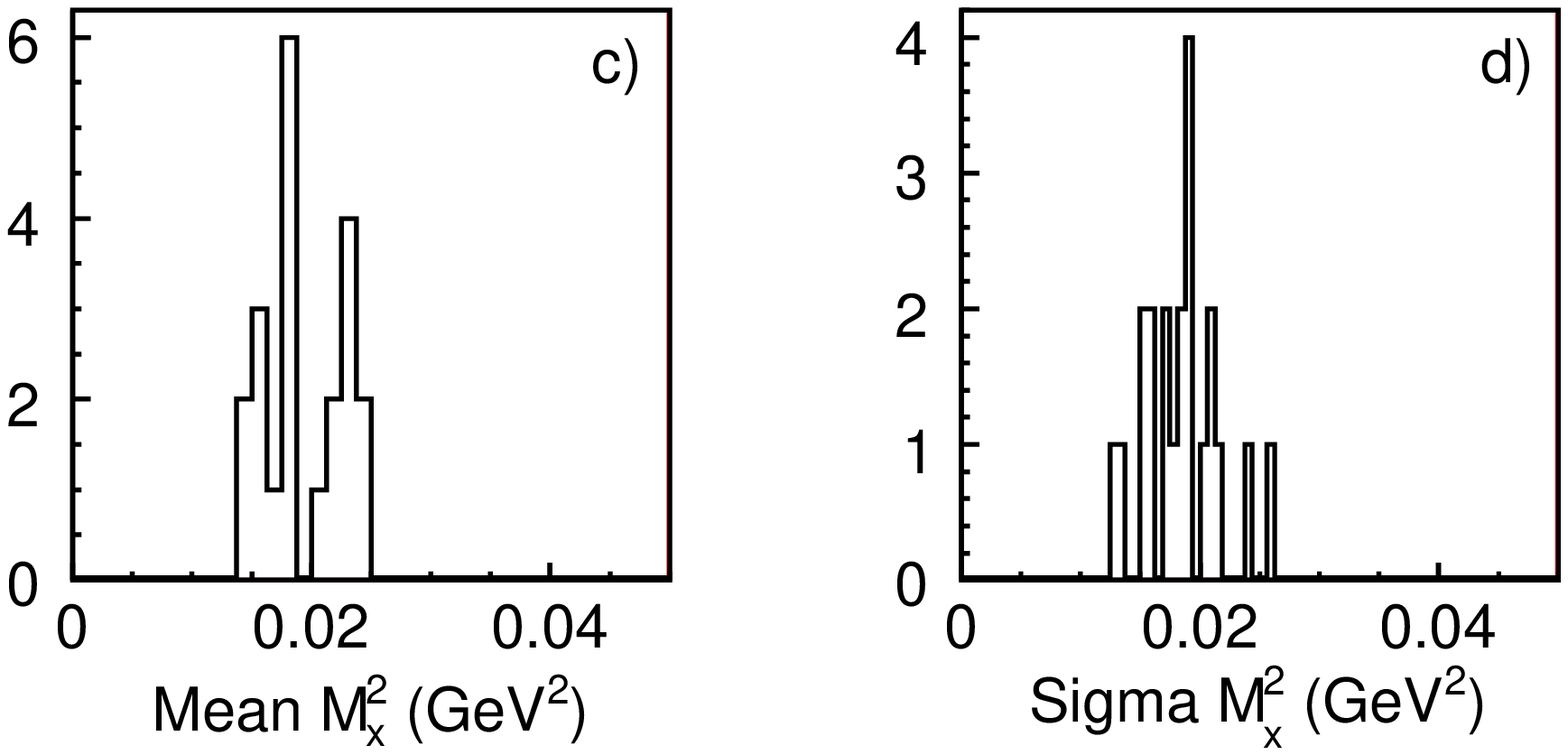}}
\caption{\label{fig:GP_PEAK_HIST} Distribution of the means and standard deviations of the Gaussian
functions for the photon and pion peaks for sub-bins of one kinematical bin:
$1.7$~(GeV/c)$^2<Q^2<2.8$~(GeV/c)$^2$ and
$0.1$~(GeV/c)$^2<|t|<0.8$~(GeV/c)$^2$}.
\end{figure*}

Additional sources of systematic uncertainties are the dependences of the
mean values and standard deviations of the functions $G_\gamma$ and
$G_{\pi^0}$ on the kinematics of an event. In
Fig.~\ref{fig:GP_PEAK_HIST}, distributions of these parameters
obtained in smaller sub-bins of $t$ and $Q^2$ 
within the $Q^2$ bin of
$1.7$ (GeV/c)$^2$ to $2.8$ (GeV/c)$^2$ and $t$ bin of
$0.1$ (GeV/c)$^2$ to $0.8$ (GeV/c)$^2$ are shown. Single photon
$[$Bethe-Heitler events, selected as shown in Fig.~\ref{fig:pxpy}$]$ events
in this $t $ and $Q^2$ bin were divided into $30$ sub-bins
(Fig.~\ref{fig:GP_PEAK_HIST}.(a,b)), while single pion events were
divided into $21$ sub-bins (Fig.~\ref{fig:GP_PEAK_HIST}(c,d)). Physics asymmetries in each $t$
and $Q^2$ bin were extracted using the central values of these
parameters in the given bin. To estimate the systematic uncertainties, the whole fitting
procedure was repeated with additional four values of these parameters
within a $\pm 1$ RMS range. Thus, a total of $5^4=625$
different sets of means and standard deviations were tested for each
$t$ and $Q^2$ bin.

\begin{figure}[t!]
\vspace{160mm}
{\includegraphics{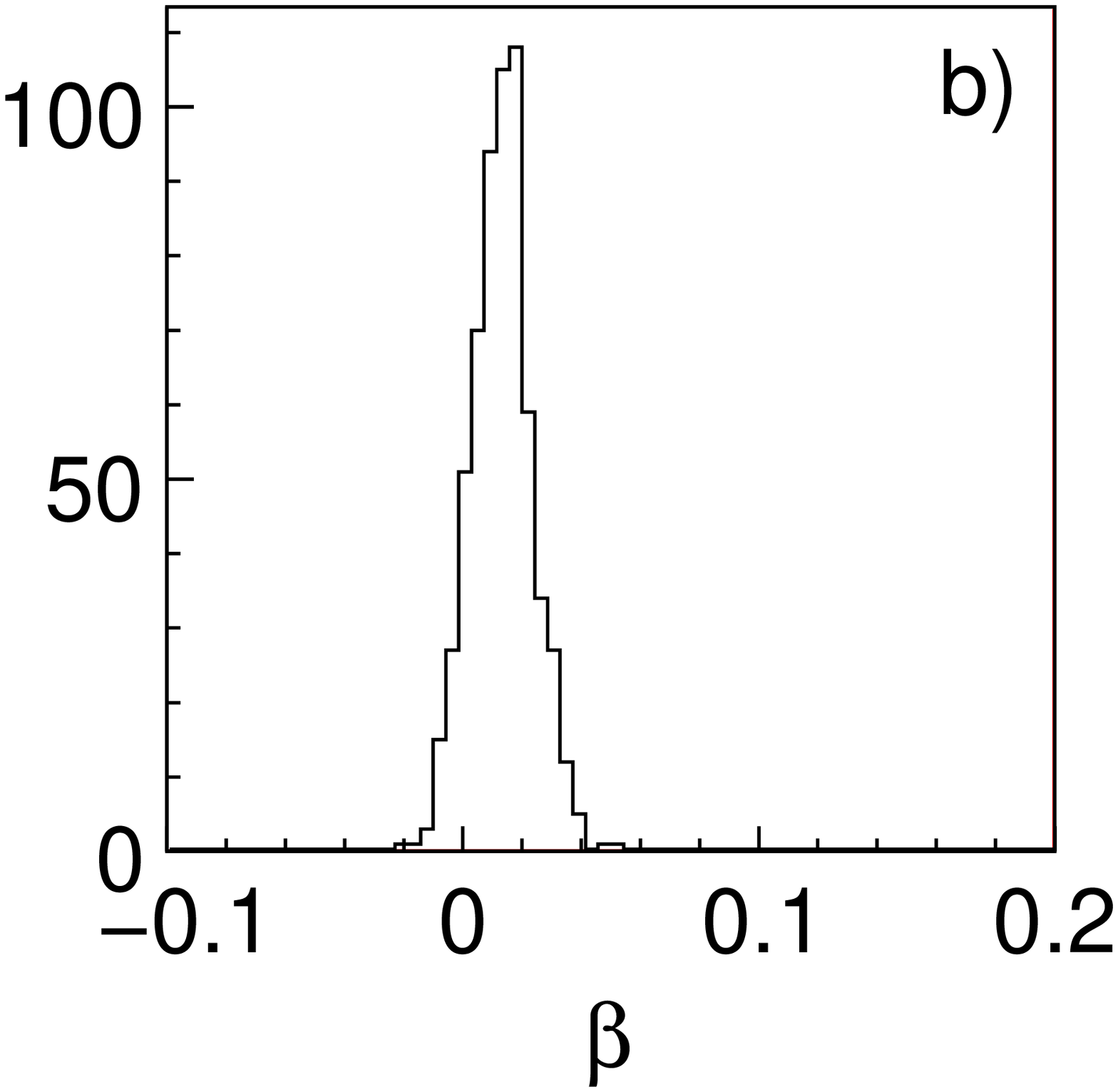}}
{\includegraphics{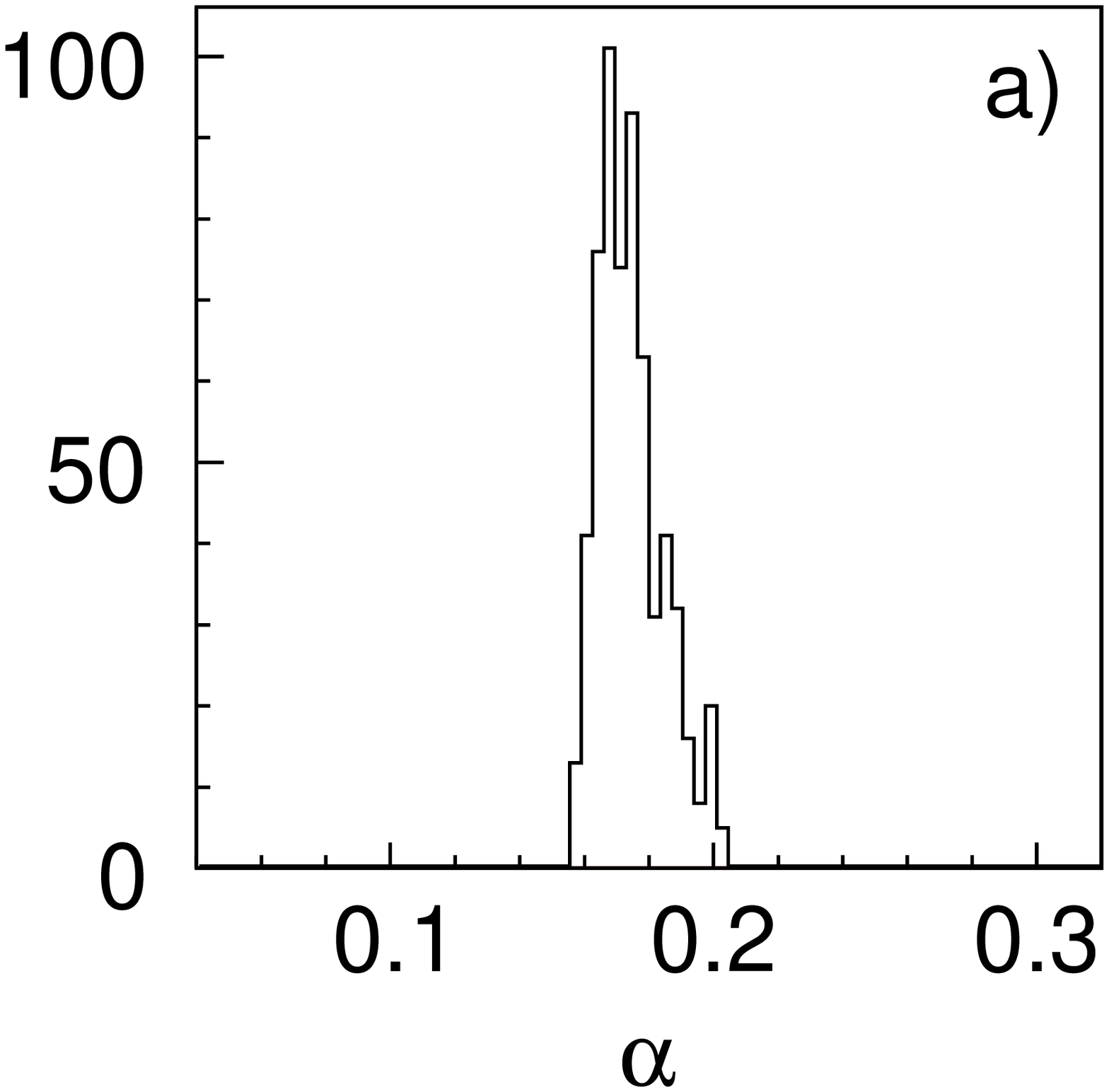}}
\caption{\label{fig:SYSTEMATIC_ERR} Distribution of the $\sin{\phi_{p\gamma}}$
(a) and $\sin{2\phi_{p\gamma}}$ (b) moments of the beam spin asymmetries
for the $Q^2$ bin of $1.35$ (GeV/c)$^2$ to $1.7$ (GeV/c)$^2$ and $0.1$~(GeV/c)$^2<|t|<0.4$~(GeV/c)$^2$,
extracted from the fits 
to the missing mass squared distributions using different values of means and
standard deviations of $G_\gamma$ and $G_{\pi^0}$ within $\pm 1$ RMS. }
\end{figure}

In Fig.~\ref{fig:SYSTEMATIC_ERR}a, the distribution of the
$\sin{\phi_{p\gamma}}$ moments obtained for different sets of Gaussian
parameters is shown for the $Q^2$ and $t$ bin of $1.35$ to $1.7$
(GeV/c)$^2$ and $0.1$ to $0.4$ (GeV/c)$^2$, respectively. The
RMS of this distribution is $0.015$. The same
distribution for the $\sin{2\phi_{p\gamma}}$ moment has an RMS of $0.013$,
see Fig.~\ref{fig:SYSTEMATIC_ERR}b. These RMS values
were taken as the systematic uncertainties 
of the moments due to the kinematical dependence of the parameters of
$G_\gamma$ and $G_{\pi^0}$ within the kinematic bin.  

\clearpage

\section{Results}

\indent

\begin{figure}[ht!]
\vspace{100mm}
{\includegraphics{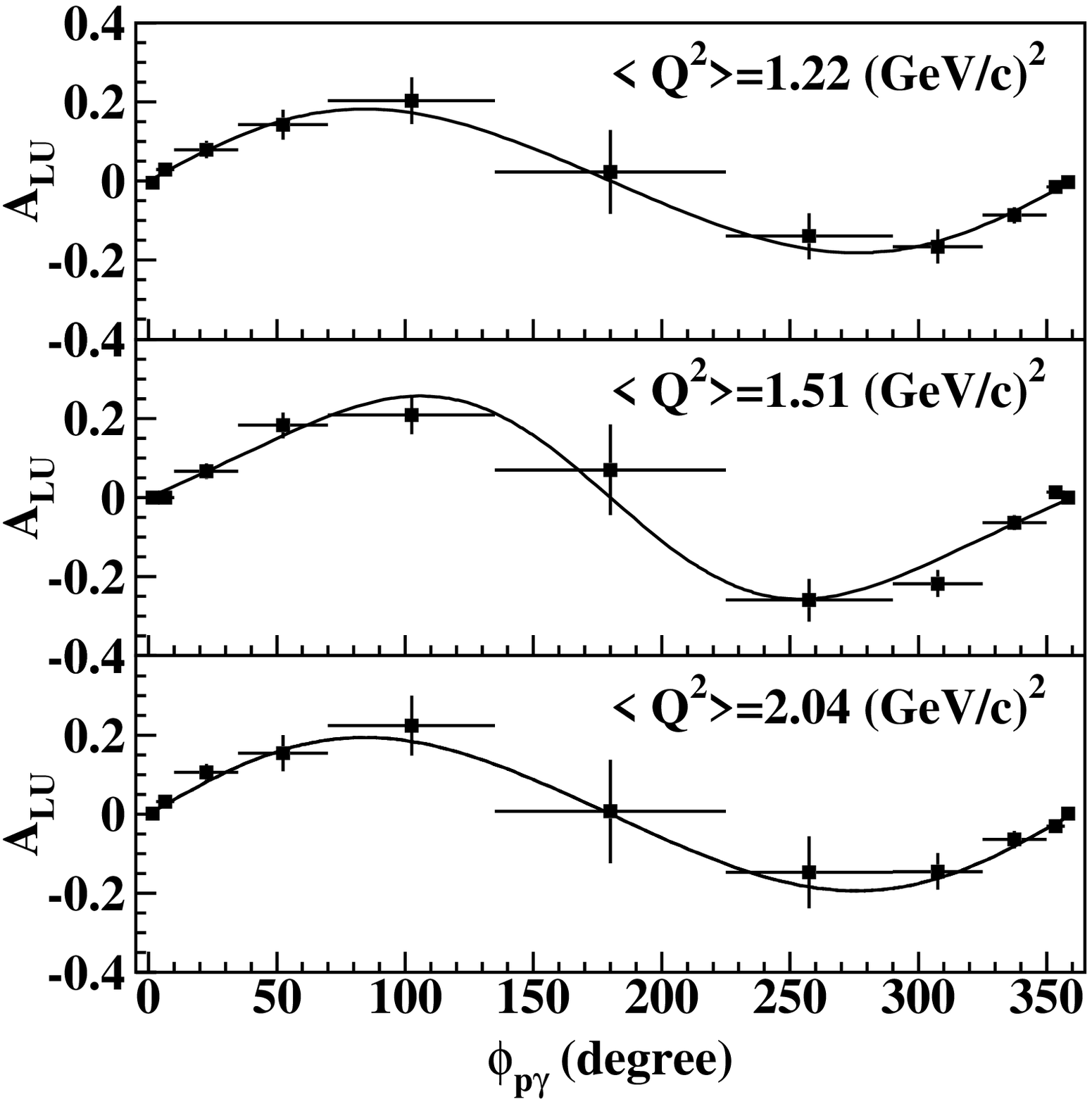}}
\caption{\label{fig:ALL_3_ASYMETRIES_Q2} Asymmetry ($A_{LU}$) as a function of
  azimuthal angle fitted according to Eq.(\ref{eq:ASYMSIN}) to extract
 the $\sin\phi$ and $\sin{2\phi}$ moments in three $Q^2$ bins, see
  Table~\ref{tab:ASYM_ALL_RESULTS_Q2}.}  
\end{figure}

\begin{figure}[ht!]
\vspace{100mm}
{\includegraphics{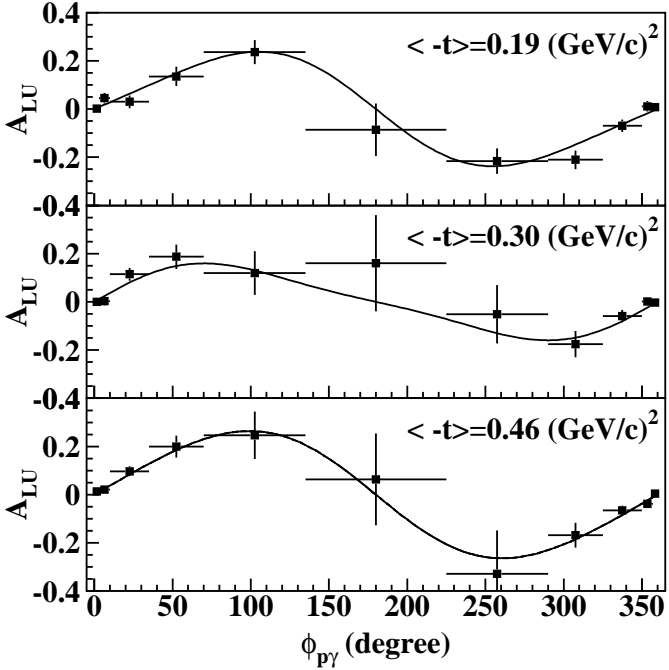}}
\caption{\label{fig:ALL_3_ASYMETRIES_t} Asymmetry as a function of
  azimuthal angle fitted according to Eq.(\ref{eq:ASYMSIN}) to extract
  the $\sin{\phi}$ and $\sin{2\phi}$ moments in three $t$ bins, see
  Table~\ref{tab:ASYM_ALL_RESULTS_t}.}  
\end{figure}

Our results on the beam spin asymmetry, calculated using
Eq.(\ref{eq:CACL_ASYM}), are presented in Tables \ref{tab:ALU_Q2}
and \ref{tab:ALU_t}. In order to extract $\sin{\phi_{p\gamma}}$
($\alpha$) moments as a function of $Q^2$ and $t$, the azimuthal angular
dependence of $A_{LU}$ was 
fitted with the function presented in Eq.(\ref{eq:ASYMSIN}).
In Fig.~\ref{fig:ALL_3_ASYMETRIES_Q2} the $\phi_{p\gamma}$ dependences of
$A_{LU}$ for three bins of $Q^2$ for $0.1$~(GeV/c)$^2<-t<0.4$~(GeV/c)$^2$ are
shown. 
The azimuthal angular dependence of $A_{LU}$ for three bins of $t$ at
$1.4$~(GeV/c)$^2<Q^2<2.5$~(GeV/c)$^2$ 
 is shown in Fig.~\ref{fig:ALL_3_ASYMETRIES_t}.
The solid line is a fit of the extracted asymmetry using the function 
 presented in Eq.(\ref{eq:ASYMSIN}). 
In both figures only statistical uncertainties are shown. 

\begin{table}[ht!]
\begin{center}
\begin{tabular}{|c|c|c||c|c||c|c|} 
\hline 
$<Q^2>$ &\multicolumn{2}{|c|}{$1.22$
[(GeV/c)$^2$]}&\multicolumn{2}{|c|}{$1.51$ [(GeV/c)$^2$]}&
\multicolumn{2}{|c|}{$2.04$ [(GeV/c)$^2$]} \\ 
\hline 
 $<\phi>$  & $A_{LU}$ & $\delta A_{LU}$ &$A_{LU}$ & $\delta A_{LU}$ &$A_{LU}$ & $\delta A_{LU}$  \\ 
\hline 
   1.5$^\circ$ &-0.0051 & 0.0056 & 0.0007 & 0.0055&  0.0021 & 0.0060 \\
   6.5$^\circ$ & 0.0289 & 0.0139 &-0.0005 & 0.0134&  0.0315 & 0.0154 \\
  22.5$^\circ$ & 0.0794 & 0.0221 & 0.0660 & 0.0196&  0.1054 & 0.0217 \\
  52.5$^\circ$ & 0.1427 & 0.0381 & 0.1826 & 0.0322 & 0.1544 & 0.0454 \\
 102.5$^\circ$ & 0.2028 & 0.0589 & 0.2080 & 0.0475 & 0.2237 & 0.0750 \\
 180.0$^\circ$ & 0.0227 & 0.1062 & 0.0703 & 0.1146 & 0.0071 & 0.1307 \\
 257.5$^\circ$ &-0.1394 & 0.0582 &-0.2592 & 0.0539 &-0.1471 & 0.0906 \\
 307.5$^\circ$ &-0.1662 & 0.0431 &-0.2173 & 0.0337 &-0.1451 & 0.0463 \\
 337.5$^\circ$ &-0.0862 & 0.0203 &-0.0631 & 0.0185 &-0.0643 & 0.0220 \\
 353.5$^\circ$ &-0.0147 & 0.0135 & 0.0130 & 0.0129 &-0.0308 & 0.0158 \\
 358.5$^\circ$ &-0.0032 & 0.0063 &-0.0001 & 0.0058 & 0.0006 & 0.0063 \\
\hline 
\end{tabular} 
\end{center} 
\caption{\label{tab:ALU_Q2} Calculated beam spin
asymmetry, $A_{LU}$, at each $\phi_{p\gamma}$ for three $Q^2$ bins, see
Fig. \ref{fig:KINEMATICS_Q2_T_XB}. The average values of $x_B$ and $t$
are shown in Table~\ref{tab:ASYM_ALL_RESULTS_Q2}.} 
\end{table} 
\begin{table}[ht!]
\begin{center}
\begin{tabular}{|c|c|c||c|c||c|c|} 
\hline 
$<-t>$ &\multicolumn{2}{|c|}{$0.19$
[(GeV/c)$^2$]}&\multicolumn{2}{|c|}{$0.30$ [(GeV/c)$^2$]}&
\multicolumn{2}{|c|}{$0.46$ [(GeV/c)$^2$]} \\ 
\hline  
 $<\phi>$  & $A_{LU}$ & $\delta A_{LU}$ &$A_{LU}$ & $\delta A_{LU}$
&$A_{LU}$ & $\delta A_{LU}$  \\  
\hline 
   1.5$^\circ$ &  0.0014 & 0.0102  &-0.0002  & 0.0071  & 0.0141  & 0.0049 \\ 
   6.5$^\circ$ &  0.0449  & 0.0219 &  0.0024  & 0.0185  & 0.0209  & 0.0131 \\ 
  22.5$^\circ$ &  0.0298  & 0.0260 &  0.1143 &  0.0266  & 0.0969 &  0.0204 \\ 
  52.5$^\circ$ &  0.1354 &  0.0402 &  0.1871  & 0.0503 &  0.1998  & 0.0450 \\ 
 102.5$^\circ$ &  0.2361 &  0.0503  & 0.1191 &  0.0906 &  0.2461 &  0.0984 \\ 
 180.0$^\circ$ & -0.0858 &  0.1092  & 0.1608 &  0.1994  & 0.0629  & 0.1908 \\ 
 257.5$^\circ$ & -0.2164  & 0.0533  &-0.0513 &  0.1210 & -0.3280 &  0.1797 \\ 
 307.5$^\circ$ & -0.2105 &  0.0389  &-0.1756 &  0.0548  &-0.1679  & 0.0515 \\ 
 337.5$^\circ$ & -0.0691 &  0.0253  &-0.0594 &  0.0260 & -0.0653  & 0.0194 \\ 
 353.5$^\circ$ &  0.0111 &  0.0218 &  0.0010 &  0.0179 & -0.0377  & 0.0129 \\ 
 358.5$^\circ$ &  0.0075 &  0.0110 & -0.0035 &  0.0076 &  0.0045 &  0.0053 \\ 
\hline 
\end{tabular} 
\end{center} 
\caption{\label{tab:ALU_t} Calculated beam spin
asymmetry, $A_{LU}$, at each $\phi_{p\gamma}$ for three $t$ bins, see
Fig. \ref{fig:KINEMATICS_Q2_T_XB}. The
average values of $x_B$ and $Q^2$ are shown in
Table~\ref{tab:ASYM_ALL_RESULTS_t}.} 
\end{table} 

The systematic uncertainties on $\alpha$ have been evaluated using the studies presented in the
previous section. Besides the systematic uncertainties due to the $M_x^2$ fits, $3\%$, and
the determination of the parameters of $G_\gamma$ and $G_{\pi^0}$,
$\pm 8\%$, there is an uncertainty in the calculation of the $A_{LU}$ from
the determination of the beam polarization, $P_e$, and an uncertainty due to the
charge asymmetry. The uncertainty on the beam polarization was estimated
to be $\pm 3\%$. The helicity-dependent charge asymmetry was about
$0.7\%$. The estimated total systematic uncertainties for the 
extracted moments are $-8.7\%$ and $+8.2\%$.  

\begin{table*}[ht!]
\begin{center}
\begin{tabular}{|c|c|c||c|c|c|c|c|} 
\hline 
 $<Q^2>$~[(GeV/c)$^2$]  & $<x_B>$ & $<-t>$~[(GeV/c)$^2$] & $ \alpha $ & $ \beta$ &$\alpha^\prime$ &$\gamma$ \\ 
\hline 
\hline 
1.22  & 0.17 & 0.23 &  $ 0.181 \pm  0.032$&  $ 0.099 \pm  0.023$ &$0.181\pm 0.032$& $-0.098\pm 0.228$ \\ 
1.51 & 0.20 & 0.26 &  $ 0.245 \pm  0.028$ &  $-0.040 \pm  0.021$ &$0.234\pm 0.024$& $0.319 \pm 0.195$ \\ 
2.04 & 0.28  & 0.38 &  $ 0.192 \pm  0.044$ &  $ 0.010 \pm  0.030$& $0.191\pm 0.045$& $-0.107 \pm 0.288$ \\ 
 \hline 
\end{tabular} 
\end{center} 
\caption{\label{tab:ASYM_ALL_RESULTS_Q2} Results from the fits to the $\phi$ dependences of $A_{LU}$ with the functions presented in Eqs. (\ref{eq:ASYMSINp}) and (\ref{eq:ASYMSIN}). Only statistical uncertainties are presented.}
\end{table*} 

\begin{table*}[ht!]
\begin{center}
\begin{tabular}{|c|c|c||c|c|c|c|c|} 
\hline 
 $<-t>$~[(GeV/c)$^2$] & $<x_B>$ & $<Q^2>$~[(GeV/c)$^2$] & $ \alpha $ & $ \beta $&$\alpha^\prime$&$\gamma$  \\ 
\hline 
\hline 
$0.19$ & $0.23$ & $1.69$ &  $ 0.228 \pm  0.029$ &  $ -0.034 \pm  0.024$ 
&$0.222\pm 0.026$& $0.318 \pm 0.248$ 
\\ 
 $0.30$ & $0.23$ & $1.74$ &  $ 0.147 \pm  0.055$ &  $ 0.033 \pm  0.037$ 
&$0.161\pm 0.045$& $-0.223 \pm 0.274$ 
\\ 
 $0.46$ & $0.25$ & $1.83$ &  $ 0.260 \pm  0.062$ &  $ -0.022 \pm  0.039$ 
&$ 0.260\pm 0.057$& $0.200 \pm 0.331$ 
\\ 
\hline 
\end{tabular} 
\end{center} 
\caption{\label{tab:ASYM_ALL_RESULTS_t}  Results from the fits to the $\phi$ dependences of $A_{LU}$ with the functions presented in Eqs. (\ref{eq:ASYMSINp}) and (\ref{eq:ASYMSIN}). Only statistical uncertainties are presented.}
\end{table*}

The final results on the moments $\alpha$ and $\beta$
are presented in
Table~\ref{tab:ASYM_ALL_RESULTS_Q2} and
Table~\ref{tab:ASYM_ALL_RESULTS_t}. For consistency checks, the $\phi_{p\gamma}$
dependence of the $A_{LU}$ were fitted with the function presented in
Eq.(\ref{eq:ASYMSINp}). The obtained results on $\alpha^\prime$ and
$\gamma$ are also presented in Table~\ref{tab:ASYM_ALL_RESULTS_Q2} and
Table~\ref{tab:ASYM_ALL_RESULTS_t}. 
In the tables only the statistical uncertainties are quoted. 

\begin{figure*}[ht]
\vspace{100mm}
{\includegraphics{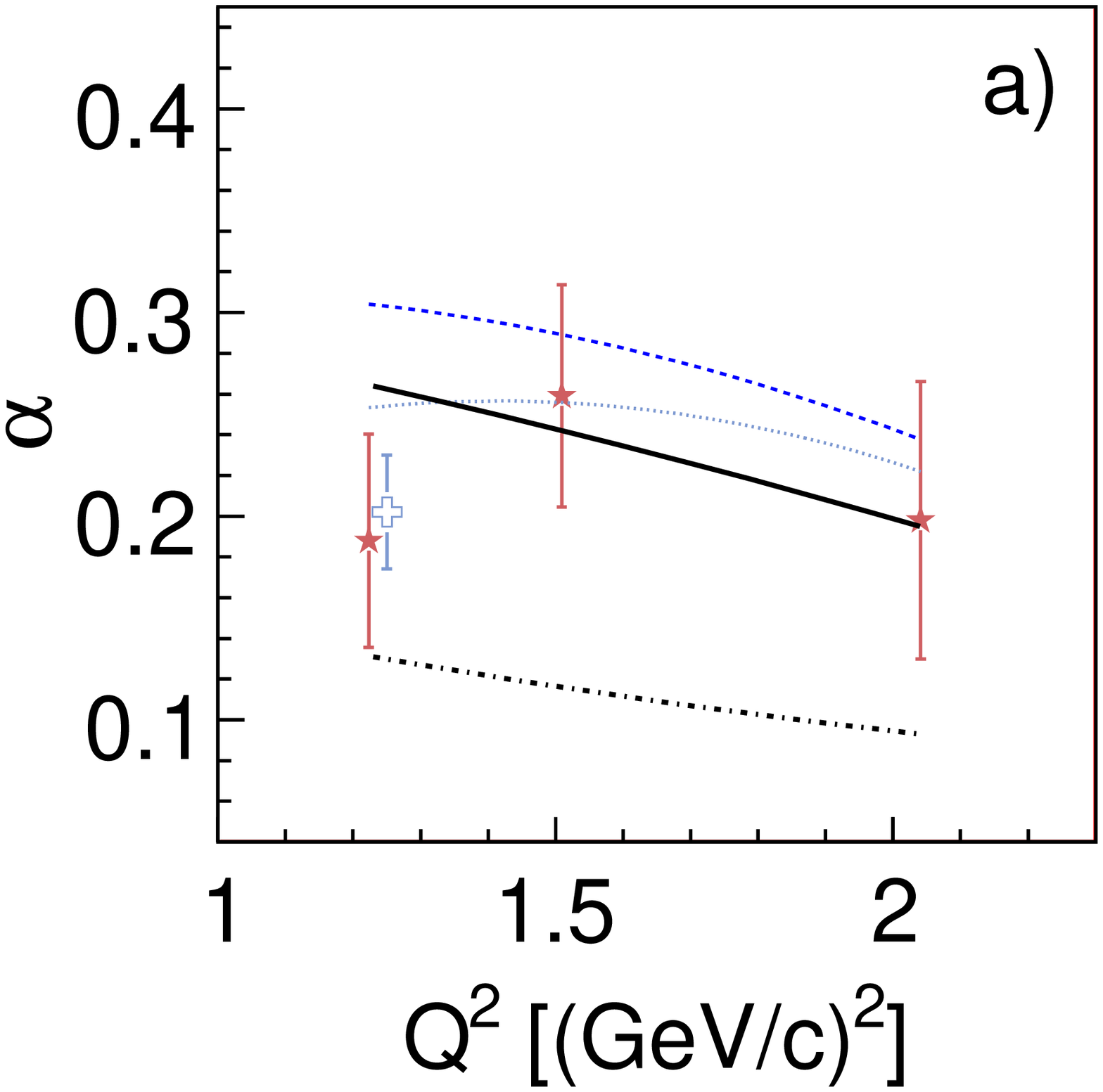}}
{\includegraphics{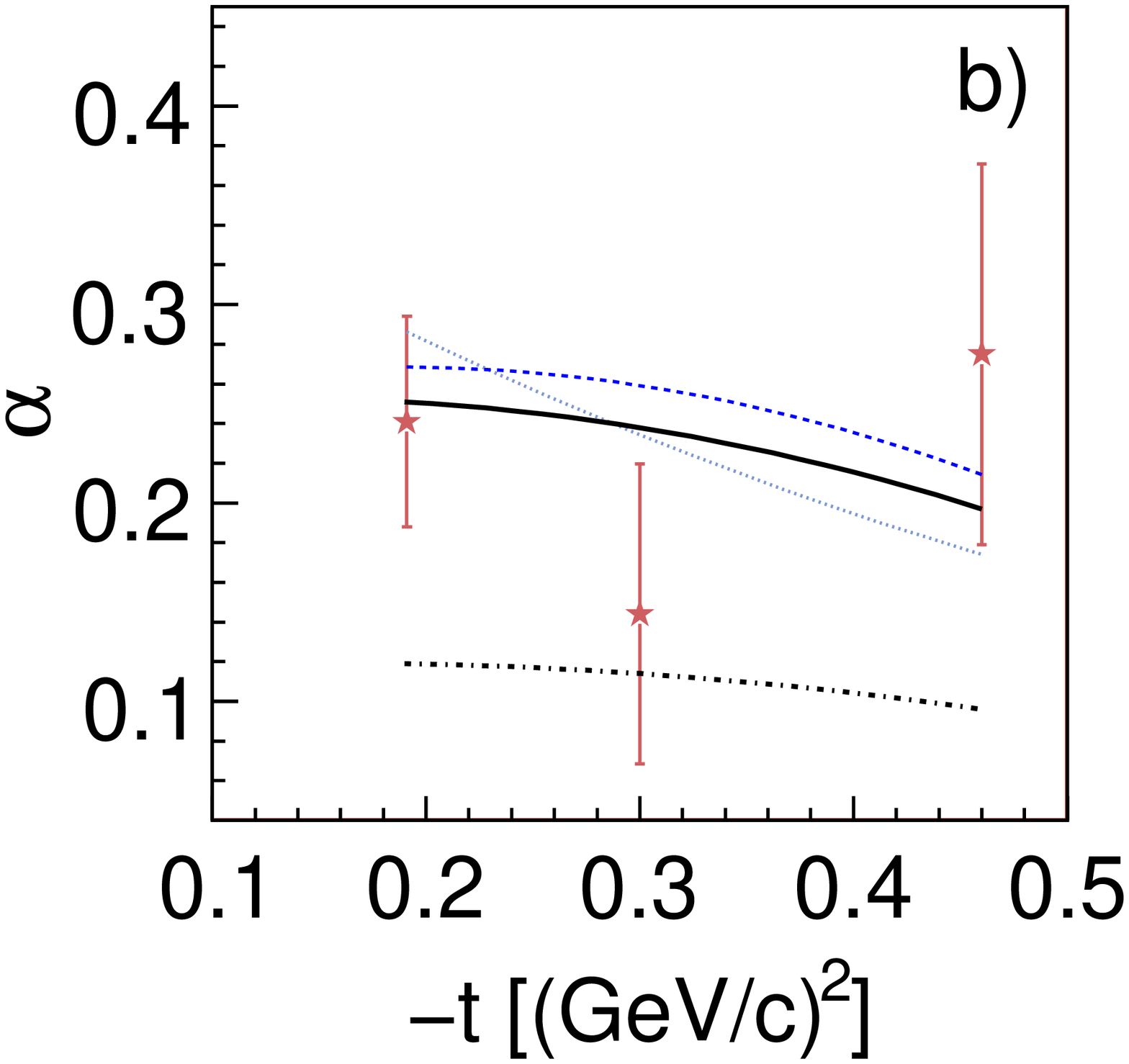}}
\caption{\label{fig:WORLD_DATA_DVCS2} (Color) The $Q^2$ (a) and $t$
(b) dependences of the $\sin{\phi_{p\gamma}}$ moments of the DVCS
  asymmetry. The ``cross'' symbol is 
the asymmetry measured by CLAS with the $4.2$ GeV data.
The curves are model
calculations: the solid and dashed-dotted lines are the calculations
based on a Regge model for the photon-proton interaction \cite{Laget}. The
dashed and dotted
lines are from the GPD-based model~\cite{GVG1,GPV}. Each curve represents
different parameter settings (see text for details). In the plots the error bars are the combined statistical
and systematical uncertainties added in quadrature.}  
\end{figure*}

In Fig.~\ref{fig:WORLD_DATA_DVCS2}, the $Q^2$ and $t$-dependences of
the $\sin{\phi_{p\gamma}}$ moment ($\alpha$) are shown. In the graph for
the $Q^2$ dependence the previously 
published CLAS result on $\alpha$ \cite{Clas01a} is shown by an open cross.
The solid and dashed-dotted lines are the $Q^2$ and
$t$-dependences of the $\sin{\phi_{p\gamma}}$ moments of the beam spin asymmetry
calculated using a Regge trajectory exchange model for the 
interaction of the photon with the proton \cite{Laget03,Laget} 
using two treatments of the unitarity cuts. The dashed-dotted
curves retain the contribution of the poles \cite{Laget03} and the
contribution of the $\rho$-$p$ elastic cut \cite{Laget}. They take into
account the coupling between the $\gamma$-$p$ and the $\rho$-$p$ channels
and are well under control since they rely on known on-shell matrix
elements. The solid curves also take into account all of the
inelastic diffractive cuts, under the assumption that diffractive dissociation
saturates the photo-absorption cross section. This treatment leads to
an excellent agreement with the 
unpolarized cross sections, as well as the helicity difference of the DVCS cross sections,
recently measured in Hall A \cite{halla}.

The other two curves in the graphs represent calculations of the
$\sin{\phi_{p\gamma}}$ moments with a GPD-based model 
\cite{GVG1}, using a Regge {\it ansatz} for the
$t$-distribution of the GPD \cite{GPRV}. The parameterization of
GPDs includes the D-term (to ensure the polynomiality of Mellin
moments of GPDs \cite{radd}). Helicity-dependent
cross sections were calculated at the twist-three level and include target mass
corrections \cite{GVG1}. Calculations are done for two different skewedness
parameter values for sea quarks, $b_{sea}=1$ - dashed line, and $b_{sea}=2$ -
dotted line. In the calculations the profile parameter for the valence
quark was $b_{val} = 1$ and the Regge slope parameter $\alpha^\prime =
1.05$ \cite{GPRV}. The data favor the GPD model with the sea quark
skewedness parameter $b_{sea}=2$. The calculations
include the DVCS contributions to both the numerator
(Eq.(\ref{InterferenceTerm})) and
denominator (Eq.(\ref{AmplitudesSquared})) of $A_{LU}$. 

In the method used for separation of the single photon events, the
radiative tails were folded into the ``background''. The radiative
effects are inseparable from the Born level \Epg~ process. Detailed
radiative calculations by Vanderhaeghen et al. \cite{qedrad} showed
that the effect on the beam spin asymmetry is less than $5\%$. This is
much smaller than the uncertainties of our measurements.

The extracted $\sin{2\phi_{p\gamma}}$ moment or  $\cos{\phi_{p\gamma}}$ denominator term of $A_{LU}$ 
is within uncertainties consistent with zero. 
The precision of
our results is not good enough to determine small contributions
from higher twists and the power suppressed terms. 

\section{Summary}

\indent

The beam spin asymmetry $A_{LU}$ in the deeply virtual
production of real photons has been measured
using a $4.8$ GeV longitudinally polarized electron beam and the CLAS
detector. 
The experimental $A_{LU}$ for each kinematical bin
was calculated using the number of single photon events extracted 
from the fits to the line shape of the missing mass squared ($M_x^2$)
distributions. Studies using Bethe-Heitler and $\pi^0$
production events have been performed to test the validity of the fit
method. It was found that the systematic uncertainties on the moments from the fit
procedure are not more than $6\%$. Overall systematic uncertainties have been
estimated to be less than $9\%$ of the value of $\alpha$.

The $\sin{\phi_{p\gamma}}$, $\sin{2\phi_{p\gamma}}$, and $\cos{\phi_{p\gamma}}$ moments of the
azimuthal angular dependence of the asymmetry have been extracted in three
bins of $Q^2$ and three bins of transferred momentum $t$ (see
Tables~\ref{tab:ASYM_ALL_RESULTS_Q2} and
\ref{tab:ASYM_ALL_RESULTS_t}). The $Q^2$ 
and $t$ dependences of the $\sin{\phi_{p\gamma}}$ moment are
compared with theoretical calculations using the GPD-based model for
DVCS \cite{GVG1,GPV} and the photon-proton interaction based on a Regge model
\cite{Laget}.  The sensitivity of
our results is not good enough to estimate small contributions
from higher twists and the power suppressed terms. However, the data
put constraints on the model parameters used to calculate the beam spin
asymmetry. Also, the data are reproduced by a model based on Regge
poles and unitary cuts. Whether this is a consequence of the quark
hadron duality remains to be investigated.
Clearly, more data in a wide range of kinematics are needed to
refine the parameters of the models. New experiments on DVCS
and on Deeply Virtual Meson Production will provide 
further observables in the Deeply Exclusive Production regime that
can be used in a global fit to extract the GPDs.

\section{Acknowledgments}

We would like to acknowledge the outstanding work of the staff of the
Accelerator Division and the Physics
Divisions, and the Hall B technical staff that made this experiment possible. 
We would also like to acknowledge useful discussions with
M. Vanderhaeghen and for providing a computer code for model calculations. 

This work was supported in part by the Istituto Nazionale di Fisica 
Nucleare, the French Centre National de la Recherche Scientifique, the French Agence Natioanale de la Recherche, 
the French Commissariat \`{a} l'Energie Atomique, the U.S. Department of 
Energy, the National 
Science Foundation and the Korea Research Foundation.
Authored by The Southeastern Universities Research Association,
Inc. under U.S. DOE Contract No. DE-AC05-84150. The U.S. Government
retains a non-exclusive, paid-up, irrevocable, world-wide license to
publish or reproduce this manuscript for U.S. Government purposes.

\end{document}